\documentclass[a4paper,11pt]{article}
\usepackage{jheppub}

\usepackage{array} 
\usepackage{derivative}
\usepackage{multirow}
\usepackage{tikz}
\usepackage{subcaption}
\usepackage{colorblind}
\usepackage{dsfont}
\usepackage{mathtools}
\usepackage{braket}
\usepackage{bm}
\usepackage{booktabs}
\usepackage{float}

\newcolumntype{C}[1]{>{\arraybackslash}p{#1}}
\newcommand{\eps}{\epsilon}

\newcommand{\dd}{\mathrm{d}}

\newcommand{\G}{\mathrm{G}}
\title{Heavy-quark initiated charged-current deep-inelastic scattering coefficient functions
through $\bm{\mathcal{O}(\alpha_s^2)}$}
\author{Kirill Kudashkin}
\affiliation{School of Physics, Shandong University, Jinan, Shandong 
250100, China}
\affiliation{Tif Lab, Dipartimento di Fisica, Universit`a di Milano and INFN, Sezione di Milano, Via Celoria 16, I-20133 Milano, Italy}

\emailAdd{kkudashkin@sdu.edu.cn}

\abstract{We compute the coefficient functions for heavy-quark initiated charged-current 
deep-inelastic scattering through $\mathcal{O}(\alpha_s^2)$, retaining full mass dependence 
for a single heavy-quark flavor. The calculation employs a cut-based approach to isolate 
individual diagram contributions and to derive differential equations for the relevant master 
integrals, which are solved analytically in terms of generalized polylogarithms. The results 
are presented in the decoupling scheme for $n_L$ light active flavors, facilitating their 
direct implementation in variable-flavor number schemes such as ACOT and FONLL.}
\begin{document}
	
\maketitle
\flushbottom

\section{Introduction}
The role of heavy quarks, specifically the charm quark, in the proton structure has a long  
history in theoretical physics. It began with the work of Witten~\cite{Witten:1975bh}, who 
established that charm quarks and antiquarks inside the proton are generated primarily via the 
dynamical mechanism of perturbative Quantum Chromodynamics (pQCD), known as \emph{extrinsic} 
production. Shortly thereafter, Brodsky et al.~\cite{Brodsky:1980pb} proposed that the proton 
wave function could contain a non‑perturbative \emph{intrinsic} charm component arising from 
quantum fluctuations. Deep‑inelastic scattering (DIS) at high‑energy colliders offers the 
essential experimental framework to resolve and study both contributions 
\cite{Blumlein:2015qcn}.

At present, parton distribution functions (PDFs), describing the dynamics of quarks 
and gluons inside hadrons, are extracted through a global QCD fit that combines data spanning a 
wide energy range from various experiments (CTEQ \cite{Hou:2019efy,Ablat:2024muy}, MSHT 
\cite{Harland-Lang:2014zoa,Bailey:2020ooq}, NNPDF \cite{NNPDF:2017mvq,NNPDF:2021uiq}, and 
PDF4LHC \cite{PDF4LHCWorkingGroup:2022cjn}). A dedicated theoretical framework is required to 
incorporate heavy-quark mass effects consistently across this wide energy range. 
Building on the foundation laid by~\cite{Witten:1975bh,Brodsky:1980pb}, 
variable-flavor number schemes (VFNS) such as ACOT~\cite{Aivazis:1993kh, Aivazis:1993pi} (and 
its modifications from e.g.~\cite{Stavreva:2012bs,Risse:2025smp}) and 
FONLL~\cite{Cacciari:1998it,Forte:2010ta,Ball:2015dpa}, provide such framework by implementing 
perturbative matching conditions between pQCD calculations with different numbers of active 
quark flavors.

The purpose of a VFNS is to interpolate between two characteristic kinematic regimes: 
the high‑energy regime $Q^2/m_{\mathrm{Q}}^2 \gg 1$ ($Q^2$ being the typical pQCD scale, 
$m_{\mathrm{Q}}^2$ the heavy-quark mass squared), where large logarithms of this ratio 
dominate, and the threshold region $Q^2 \approx m_{\mathrm{Q}}^2$, where finite‑mass effects 
are essential \cite{Aivazis:1993kh,Aivazis:1993pi,Forte:2010ta}. The interpolation is realized 
through universal PDF matching conditions that are most directly 
obtainable through two complementary approaches: via the operator product 
expansion \cite{Bierenbaum:2007qe}, or by explicit calculation using DIS 
coefficient functions as the suitable observable \cite{Buza:1995ie}. Accordingly, the accuracy 
of a global QCD fit in a VFNS relies on the availability of high-order pQCD calculations. The 
charged-current DIS coefficient functions computed here provide precisely this necessary input, 
from which, in principle, the matching conditions can be subsequently extracted and the VFNS 
implemented.

Such calculations have now reached $\mathcal{O}(\alpha_s^3)$ accuracy, corresponding to 
third-order (N$^3$LO) corrections to the leading QCD contribution for the dominant extrinsic 
production channel in DIS-related observables \cite{Ablinger:2020snj, 
Behring:2021asx, Ablinger:2022wbb, Ablinger:2023ahe, Ablinger:2024xtt, Ablinger:2025joi, 
Ablinger:2025nnq, Behring:2025avs}. In contrast, pQCD calculations relevant for constraining 
the intrinsic charm component of PDFs have seen comparatively little advancement over the past 
two decades --- the state of the art for this channel remains
$\mathcal{O}(\alpha_s)$ (NLO)~\cite{Kretzer:1998ju, Spezzano:2025kvi, Spezzano:2025rfy}. The 
primary reason for this is the small integrated 
momentum fraction carried by intrinsic charm, estimated to be only about $1$–$2\%$ of the total 
proton momentum~\cite{Brodsky:1980pb,Pumplin:2007wg}. Consequently, global QCD 
fits, more often than not, adopt the simplifying ansatz of a radiatively generated charm 
sea, explicitly excluding any intrinsic contribution. 

``However, nature does not have to subscribe to this scenario'' \cite{Pumplin:2007wg}. A recent 
NNPDF global analysis revisited the intrinsic charm hypothesis \cite{Ball:2022qks}, presenting 
evidence for its existence. To move from evidence to a definitive conclusion, improvements in 
experimental and theoretical directions are mandatory. 

Several experimental programs directly probe the proton's charm content or 
provide complementary constraints. At the LHC, dedicated forward neutrino experiments like 
FASER$\nu$ \cite{FASER:2023zcr} and SND@LHC \cite{SNDLHC:2023pun} will provide up to one 
million new DIS events by the end of the HL-LHC. They will cover a larger kinematic range than 
previous experiments and are projected to significantly constrain PDF uncertainties 
\cite{Cruz-Martinez:2023sdv}. Furthermore, a potentially powerful probe of 
intrinsic charms arises from the study of atmospheric neutrino fluxes. It has been shown 
recently~\cite{Das:2025snq} that while including an intrinsic charm component does not 
fully reconcile existing discrepancies between IceCube measurements \cite{IceCube:2013low} and 
theoretical predictions for the muon neutrino flux, it identifies the high-energy atmospheric 
neutrino spectrum as a sensitive and largely unexplored observable for this kind of physics. 
For other experimental avenues, we refer to the dedicated study \cite{Risse:2025smp}, which 
estimates the sensitivity of different DIS experiments to heavy-quark mass effects.

To benefit from the forthcoming data, the high-energy community is actively advancing toward 
N\textsuperscript{3}LO global QCD fits~\cite{Ablat:2024muy,Cridge:2024icl,NNPDF:2024nan}. In 
light of this goal, the stark imbalance between pQCD calculations for intrinsic charm 
(currently $\mathcal{O}(\alpha_s)$) and extrinsic charm (now at $\mathcal{O}(\alpha_s^3)$) 
represents a critical bottleneck. Without $\mathcal{O}(\alpha_s^2)$ correction for the 
intrinsic channel, its associated theoretical uncertainty remains uncontrolled at the precision 
targeted by N\textsuperscript{3}LO fits, potentially biasing the extraction or exclusion of an 
intrinsic charm component. We address this issue directly by providing the first calculation of 
the $\mathcal{O}(\alpha_s^2)$ corrections for heavy-quark initiated charged-current DIS, 
thereby providing essential NNLO corrections for a consistent theoretical treatment of the 
intrinsic charm component. 

The paper is organized as follows. Sec.~\ref{sec:theorySetup} lays out the theoretical
framework for heavy quark DIS, defining our notation and detailing the cut-based approach used
to isolate the relevant contributions to the coefficient functions. The core computational
machinery is presented in Sec.~\ref{sec:diffEqsCompute}, which describes the derivation of
differential equations for the master integrals, their transformation to canonical form, and
their subsequent integration for both real-radiation and virtual contributions. At the end of
this section, we cover the determination of integration constants via the PSLQ algorithm.
Sec.~\ref{sec:results} presents the renormalized coefficient functions, discussing the
renormalization scheme, the regularization of soft singularities, and providing explicit
expressions for the Born, $\mathcal{O}(\alpha_s)$, and $\mathcal{O}(\alpha_s^2)$
coefficient functions, along with details on the auxiliary files and validation. A summary and
outlook are given in Sec.~\ref{sec:conclusions}. Technical details on kinematics, projectors,
and $\gamma_5$ schemes are collected in App.~\ref{app:A}, while materials related to the master
integrals are provided in App.~\ref{app:B}.
\section{Theoretical setup for heavy-quark DIS}\label{sec:theorySetup}
In this section, we provide the technical details for the computation of corrections to the 
partonic DIS structure functions up to and including $\mathcal{O}(\alpha_s^2)$. First, we 
present the formulae relevant to all specified perturbative orders, including the normalization 
of the coefficient functions, tensor decomposition, and etc. We then introduce the 
computational techniques, with a particular focus on the application of cut-based methods to 
the computation of scattering amplitudes.
\subsection{Formalism and conventions}\label{sec:notation}
We consider deep-inelastic scattering mediated by a charged-current, specifically by the 
exchange of a virtual $\mathrm{W}^{-}(q)$ boson with a nucleon $\mathrm{P}$,
\begin{equation}\label{eq:hadronReaction}
	\mathrm{W}^{-}(q) + \mathrm{P}(P) \to \mathrm{X},
\end{equation}
where $\mathrm{X}$ denotes the inclusive hadronic final state. In the following, we focus 
solely on the hadronic tensor, leaving the leptonic part out of the discussion. We refer to \cite{Aivazis:1993kh} for 
further details.

The hadronic tensor for the process in Eq.~\eqref{eq:hadronReaction} is defined in the standard 
manner as
\begin{equation}
	W^{\mu\nu} = \frac{1}{4\pi} \sum_{\mathrm{X}}
	\mathop{\overline{\sum}}\limits_{\mathrm{spin}}
	\bra{\mathrm{P}} J^{\mu} \ket{\mathrm{X}}
	\bra{\mathrm{X}} (J^{\nu})^{\dagger} \ket{\mathrm{P}}
	(2\pi)^4 \delta^{(4)}(P + q - p_{\mathrm{X}}),
\end{equation}
where $J^{\mu}$ is the charged weak current and the sum runs over all hadronic final states 
$\mathrm{X}$ with momentum $p_{\mathrm{X}}$.

In the Bjorken limit, the dominant contribution to the hadronic tensor assumes a factorized 
form ~\cite{Aivazis:1993kh,Aivazis:1993pi},
\begin{equation}\label{eq:ansatzFactorization}
	W^{\mu\nu}(x_B, Q^2, m^2) =
	\sum\limits_{n}
	\int_{\eta}^{1} \frac{\dd \xi}{\xi}
	f_{n}(\xi, \mu^2)\,
	\widehat{\omega}_{n}^{\mu\nu}
	\left(\frac{\eta}{\xi}, \mu^2, Q^2, m^2 \right) 
	+\mathcal{O}\left(\frac{\Lambda}{Q}\right),
\end{equation}
where the sum runs over all parton species. $\Lambda$ denotes a generic
nonperturbative QCD scale, $\eta = \eta(x_B)$ is the generalized Bjorken variable defined in 
Eq.~\eqref{eq:genBjorkenVar}, and $f_{n}(\xi, \mu^2)$ is the parton distribution function for 
parton $n$ carrying a light‑cone momentum fraction $\xi \equiv p^{+}/P^{+}$ at factorization 
scale $\mu$.

We work in the massive ACOT factorization scheme, which consistently retains heavy-quark mass
effects and whose validity is established within the framework of the QCD factorization 
theorem~\cite{Collins:1998rz}. In this framework, the factorized hadronic tensor involves 
finite, mass-factorized partonic tensors $\widehat{\omega}_n^{\mu\nu}$. In the present work, 
however, we compute the corresponding massive, ultraviolet-renormalized but 
\emph{ACOT-unsubtracted} partonic tensor $\omega_n^{\mu\nu}$ through $\mathcal{O}(\alpha_s^2)$ 
for heavy-quark initiated charged-current deep-inelastic scattering. The matching to the 
massless $\overline{\mathrm{MS}}$ scheme and the resummation of quasi-collinear large 
logarithms are deferred to a future publication~\cite{kkudashkin123:2025toappear}.

The partonic tensor, which describes the hard scattering of the virtual boson off an incoming 
parton $n$, is defined as
\begin{equation}\label{eq:strfun_partonic_def}
	\omega_n^{\mu\nu} = \sum_{\mathrm{X}_{\mathrm{QCD}}}
	\mathop{\overline{\sum}}\limits_{\mathrm{color,\;spin}}
	\bra{n} J^\mu \ket{\mathrm{X}_{\mathrm{QCD}}}
	\bra{\mathrm{X}_{\mathrm{QCD}}} (J^\nu)^{\dagger} \ket{n},
\end{equation}
where the bar indicates averaging over the initial parton's color and spin, and the sum runs 
over all QCD final states $\mathrm{X}_{\mathrm{QCD}}$ at the given perturbative order.

Its perturbative expansion in $\alpha_s = g_s^2/(4\pi)$ reads
\begin{equation}\label{eq:partonicPerturbExp}
	\omega_n^{\mu\nu} = \sum_{k=0}^{\infty} \left( \frac{\alpha_s}{2\pi} \right)^k
	\omega_n^{(k),\mu\nu},
\end{equation}
where $\omega_n^{(k),\mu\nu}$ comprises all contributions of order $\alpha_s^k$. All quantities 
in this section are understood to be renormalized; details of the renormalization scheme choice 
are given in Sec.~\ref{sec:renorm}.

Specifically, we consider the process where the incoming parton is a heavy quark $\mathrm{Q}$ 
(or antiquark $\bar{\mathrm{Q}}$), identified as charm ($\mathrm{Q}\equiv c$), and the outgoing 
quark is massless\footnote{We focus on $\mathrm{W}^-$ initiated processes; the 
charge-conjugated $\mathrm{W}^+$ channels follow by symmetry, and all formulas below are given 
for the $\mathrm{W}^-$ case.}
\begin{equation}
	\mathrm{W}^-(q) + \mathrm{Q}(p)/\bar{\mathrm{Q}}(p) \to \mathrm{q}/\bar{\mathrm{q}} + 
	\mathrm{X}_{\mathrm{QCD}}.
\end{equation}
We adopt $n_F=4$ quark flavors, with $n_L=3$ light flavors (u, d, s) and $n_H=1$ heavy flavor. 
The relevant Mandelstam invariants are
\begin{equation}\label{eq:invariants}
	q^2 = -Q^2 < 0,\qquad p^2 = m^2 > 0,\qquad s = (q+p)^2.
\end{equation}

The charged weak current operator is given by:
\begin{equation}\label{eq:chargedCurrentVector}
	J^{\mu} =  \bar{\mathrm{q}}_{\alpha}\gamma^{\mu}(V-A\gamma_{5})\, 
	\mathcal{V}_{\alpha\beta}\,\mathrm{q}_{\beta},
\end{equation}
where $V$ and $A$ are the vector and axial-vector couplings, kept general for now; their 
standard values are $V = A = -\mathrm{i} g/\sqrt{2}$, with $g$ the weak coupling constant. The 
Cabibbo-Kobayashi-Maskawa (CKM) matrix $\mathcal{V}_{\alpha\beta}$ for $n_F = 4$ is
\begin{equation}\label{eq:CKMmatrixDef}
	\mathcal{V}_{\alpha\beta} =\begin{bmatrix}
		V_{\mathrm{ud}} & V_{\mathrm{us}}\\
		V_{\mathrm{cd}} & V_{\mathrm{cs}}
	\end{bmatrix},
\end{equation}
which induces the flavor decomposition of the partonic tensor
\begin{equation}\label{eq:flavorDecompostion}
	\omega^{\mu\nu}_n \equiv 
	|V_{\mathrm{ud}}|^2 \, \omega^{\mu\nu}_{n,\mathrm{ud}} + 
	|V_{\mathrm{us}}|^2 \, \omega^{\mu\nu}_{n,\mathrm{us}} + 
	|V_{\mathrm{cd}}|^2 \, \omega^{\mu\nu}_{n,\mathrm{cd}} + 
	|V_{\mathrm{cs}}|^2 \, \omega^{\mu\nu}_{n,\mathrm{cs}}.
\end{equation}
Because the DIS process we consider is initiated exclusively by a heavy quark (or antiquark) 
and we work through $\mathcal{O}(\alpha_s^2)$, the set of \emph{unique} Feynman diagrams contributing 
to the coefficient functions (Tab.~\ref{tab:intermidStates}) is obtained from the channels 
$\omega^{\mu\nu}_{n,\mathrm{cs}}$ and $\omega^{\mu\nu}_{n,\mathrm{ud}}$ alone (cf. 
Fig.~\ref{fig:typicalDiags}). To avoid 
proliferation of indices, we suppress the explicit flavor labels in the following and work 
with these flavor-stripped tensors. We will restore the appropriate CKM factors explicitly 
whenever a distinction between the individual flavor channels is required.
\begin{table}[htbp]
	\centering
	\begin{tabular}{c|c|c|c|c|c|c|c|c}
		\toprule
		&\multicolumn{1}{c|}{$\mathcal{O}(\alpha_s^0)$}& 
		\multicolumn{2}{c|}{$\mathcal{O}(\alpha_s^1)$} & 
		\multicolumn{5}{c}{$\mathcal{O}(\alpha_s^2)$} \\
		\hline
		&B & V & R & VV & RV & RRA & RRB & RRC\\
		\hline
		$\mathrm{X}_{\mathrm{QCD}}$&$\textrm{q}$ & $\textrm{q}$ & $\textrm{gq}$ & $\textrm{q}$& 
		$\textrm{gq}$ &  
		$\textrm{ggq}$, 
		$\tilde{\textrm{g}}\tilde{\textrm{g}}\textrm{q}$, $\bar{\textrm{q}}\textrm{qq}$ & 
		$\bar{\textrm{Q}}\textrm{Qq}$ & $\textrm{Q}\bar{\textrm{q}}\textrm{q}$\\
		\hline
		N&\multicolumn{1}{c|}{1}& 
		\multicolumn{2}{c|}{5} & 
		\multicolumn{5}{c}{106}\\
		\bottomrule
	\end{tabular}
	\caption{Intermediate states contributing to the coefficient 
		functions at different orders in perturbation theory. 
		Abbreviations: B (Born contribution), V (virtual contribution), R 
		(real 
		contribution), VV (double virtual contribution), RV (real-virtual 
		contribution), RRA, RRB and RRC (different real-real contributions). The bottom row 
		lists the 
		number \textrm{N} of the corresponding unique forward scattering amplitudes 
		contributing to the coefficient functions.} 
	\label{tab:intermidStates}
\end{table}

Following the normalization of Ref.~\cite{Kretzer:1998ju}, the amplitude is 
decomposed into tensors as follows:
\begin{equation}\label{eq:amplTensorStr}
	\omega^{\mu\nu} = 
	-\omega_1\eta^{\mu\nu}+\omega_2p_2^{\mu}p_2^{\nu}+
	\mathrm{i}\omega_3\varepsilon^{\mu\nu\alpha\beta}p_{1,\beta}p_{2,\alpha}+
	\omega_4p_1^{\mu}p_1^{\nu}+
	\omega_5(p_1^{\mu}p_2^{\nu}+p_2^{\mu}p_1^{\nu}),
\end{equation}
where $\eta^{\mu\nu}$ is the Minkowski metric and $\varepsilon$ is the Levi-Civita symbol. Our 
goal is to compute the NNLO QCD correction to the dominant structure functions. We omit the contributions 
of $\omega_4$ and $\omega_5$, as they are suppressed by powers of the lepton mass from the 
lepton vertex.\footnote{We note, however, that a recent NLO analysis suggests the corresponding 
structure functions can be relevant in certain cases~\cite{Spezzano:2025kvi,Spezzano:2025rfy}. 
There is no conceptual difficulty in computing $\omega_4$ and $\omega_5$ within our framework, 
but such a calculation lies beyond the original goal of the present publication.}

The form factors in Eq.~\ref{eq:amplTensorStr} are projected 
out as follows:
\begin{equation}\label{eq:extractFFs}
	\omega_{i}=\mathcal{P}_i^{\mu\nu} \omega_{\mu\nu}.
\end{equation}
Due to known technical issues with $\gamma_5$ in higher-order calculations, we employ two different schemes that address 
them: the mutual agreement of results obtained in these two schemes ensures the correctness of the underlying 
calculations. The projectors $\mathcal{P}_i$ are therefore derived in accordance with the $\gamma_5$ schemes discussed 
in App.~\ref{app:ProjANDNorm}.

Upon applying the projectors defined above to Eq.~\eqref{eq:ansatzFactorization} we 
obtain the (scaling) hadronic structure functions
\begin{equation}\label{eq:hadronStructureFunctions}
	\begin{aligned}
		F_1 & =%
		\sum_{n=Q,\bar{Q}}%
		\int_{\eta}^1%
		\frac{\dd \xi}{\xi}\,f_{n}(\xi,\mu^2)\,C_{1,n}, \\
		F_2 &=%
		\sum_{n=Q,\bar{Q}}%
		\int_{\eta}^1%
		\frac{\dd \xi}{\xi}\,f_{n}(\xi,\mu^2)\,C_{2,n}, \\
		F_3 &=%
		\sum_{n=Q,\bar{Q}}%
		\int_{\eta}^1%
		\frac{\dd \xi}{\xi}\,f_{n}(\xi,\mu^2)\,C_{3,n},
	\end{aligned}
\end{equation}
where
\begin{equation}\label{eq:DefCoefFun}
	C_{j,n} = \mathcal{N}_{j}\, \omega_{j,n} \qquad (j=1,2,3),
\end{equation}
with the normalization factors \cite{Aivazis:1993kh,Kretzer:1998ju}
\begin{equation}\label{eq:normOmega}
	\begin{aligned}
		\mathcal{N}_1 &= \frac{1}{4\pi}, \\
		\mathcal{N}_2 &= \frac{2x_B}{8\pi}\frac{\Delta^2[-Q^2,\,m^2,\,s]}{2Q^2}, \\
		\mathcal{N}_3 &= \frac{1}{4\pi}\,\Delta[-Q^2,\,m^2,\,s]. 
	\end{aligned}
\end{equation}
Here, $x_{B}$ is the Bjorken-scaling variable and $\Delta$ is the triangle function (see 
App.~\ref{app:kinematics} for definitions).

\subsection{Cut-based approach to scattering amplitudes}\label{sec:computationDetails}
\begin{figure}[!tbp]
	\centering
	\includegraphics[width=.77\textwidth]{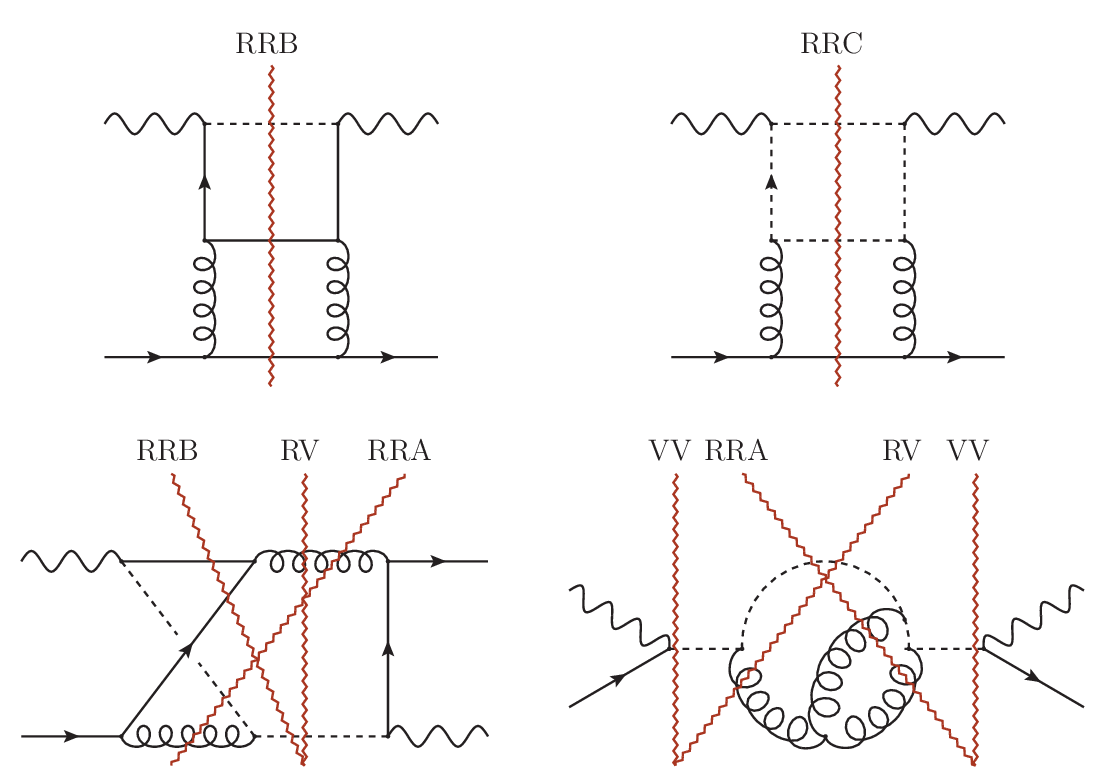}
	\caption{Typical Compton diagrams that appear in the calculations. Here, a solid line 
	indicates a heavy quark propagator, a dashed line indicates a light quark propagator, a 
	wavy line 
	indicates the external boson, and a curly line represents a gluon propagator. Zigzag lines 
	represent the cuts (cf.~Tab.~\ref{tab:cutRules}). The right diagram in the top row, and 
	similar sea diagrams, appear in 
	the coefficient functions 
	proportional to $\omega^{\mu\nu}_{n,\mathrm{ud}}$, while other diagrams are typical for 
	$\omega^{\mu\nu}_{n,\mathrm{cs}}$.}
	\label{fig:typicalDiags}
\end{figure}
The coefficient functions follow from the optical theorem, which relates the hadronic tensor 
$\omega^{\mu\nu}$ to the forward Compton amplitude \cite{Moch:1999eb}
\begin{equation}\label{eq:opticalTheorem2}
	\omega^{\mu\nu} = 2 \operatorname{Im}\mathcal{M}^{\mu\nu}.
\end{equation}
This identity allows us to compute the structure functions either by calculating 
$\mathcal{M}^{\mu\nu}$ directly or by extracting the individual contributions listed in 
Tab.~\ref{tab:intermidStates} from its imaginary part via cut techniques.\footnote{For the 
remainder of the paper, we do not distinguish between the structure functions and the forward 
scattering amplitude and use both interchangeably unless stated otherwise.}

To achieve the latter, we identify the cuts of each Feynman diagram contributing to the forward 
scattering amplitude $\mathcal{M}$ using the approach of 
Refs.~\cite{Pak:2011xt,Grigo:2014oqa,Hoff:2015kub}. In those works, the problem of finding all 
possible cuts of the given graph (Feynman diagram) is mapped onto the graph-theoretic problem 
of graph coloring. Computational algorithms for this task rely on the adjacency matrix of a 
Feynman diagram, which is straightforwardly obtained from the output of graph generators like 
\texttt{QGRAF}~\cite{Nogueira:1991ex}. Public implementations are available, e.g., in the 
\texttt{Mathematica} package \texttt{TOPOID} \cite{TopoID} and in \texttt{FeynCalc} 
\cite{Mertig:1990an,Shtabovenko:2023idz}.\footnote{We thank V. Shtabovenko for private 
communications on this matter.} For this work, we employ our own implementation of the 
algorithms described in Refs.~\cite{Pak:2011xt,Hoff:2015kub}.

We generate all graphs for the bare forward two-loop Compton amplitude using \texttt{QGRAF} and 
extract their adjacency matrices (typical two-loop diagrams are shown in 
Fig.~\ref{fig:typicalDiags}). The initial set contains graphs that possess no physical 
$s$-channel cuts (i.e., they do not contribute to the imaginary part of the forward-scattering 
amplitude). Directly filtering these graphs is cumbersome without ad‑hoc modifications in 
\texttt{QGRAF}; instead, we apply the cut‑identification algorithm described above to select 
only graphs with non‑zero $s$-channel cuts, retaining 79 out of the original 112 graphs. The 
surviving graphs and their cut information are retained for subsequent extraction of the 
contributions listed in Tab.~\ref{tab:intermidStates}.

The filtered abstract graphs are then passed to \texttt{FORM}~\cite{Vermaseren:2000nd}, where 
we apply Feynman rules (in Feynman gauge) to convert vertices 
and edges into explicit algebraic expressions. All subsequent algebraic 
manipulations are preformed in \texttt{FORM} using a combination of standard packages, notably 
\texttt{color}~\cite{vanRitbergen:1998pn} for $\mathfrak{su}(N)$ algebra, supplemented by 
private routines. These manipulations include color algebra, Dirac algebra, and operations with 
Lorentz tensors. Traces involving $\gamma_5$ are handled according to the schemes detailed in 
App.~\ref{app:ProjANDNorm}. After applying the projectors, as 
defined in Eq.~\eqref{eq:projectors} and used in Eq.~\eqref{eq:extractFFs}, the amplitude 
reduces to a linear combination of scalar Feynman integrals and rational functions of the 
kinematic variables and the space‑time dimension $d$.

We work in dimensional regularization with $d=4-2\epsilon$ to regulate both ultraviolet (UV) 
and infrared (IR) divergences that appear in the scattering amplitude.  A generic scalar 
integral in this setting has the form
\begin{equation}\label{eq:scalarIntegralDef}
	j=(\mu^2)^{2\epsilon}\int\frac{\dd^dl_1}{(2\pi)^d}%
	\frac{\dd^dl_2}{(2\pi)^d}
	\frac{\mathcal{N}(l_{1},l_{2},p_{1},p_{2})}{D_{1}^{a_1}%
		D_{2}^{a_2}%
		D_{3}^{a_3}D_{4}^{a_4}D_{5}^{a_5}D_{6}^{a_6}D_{7}^{a_7}},
\end{equation}
where $\mu$ is the 't Hooft mass, introduced to keep the coupling dimensionless in $d$ 
dimensions; if not mentioned otherwise, $\mu = Q \equiv 1$. Here, $D_i = L_i^2 - M_i^2 + 
\mathrm{i}0$ are inverse propagators; $L_i$ is a linear combination of the loop momenta 
$l_1,l_2$ and the external momenta $p_1,p_2$, while $M_i$ denotes the corresponding mass. The 
exponents $a_i$ are integers, and $\mathcal{N}(l_{1},l_{2},p_{1},p_{2})$ is a polynomial in 
scalar products of the loop and external momenta.
\begin{table}[htbp]
	\centering
	\footnotesize
	\begin{tabular}{c|l}
		\hline
		Family & Propagators $D_1, \dots, D_7$ \\
		\hline
		fam1 & $\begin{array}[t]{@{}l@{}}
			(l_{2} + p_{1})^2-m^2,\; l_{2}^2,\; (l_{1} - l_{2} - p_{1} + p_{2})^2-m^2,\\
			(l_{1} - l_{2} - p_{1})^2,\; l_{1}^2 - m^2,\; (l_{1} + p_{2})^2,\; (l_{1} - p_{1})^2
		\end{array}$ \\
		fam2 & $\begin{array}[t]{@{}l@{}}
			(l_{1} + l_{2})^2,\; l_{1}^2 - m^2,\; (l_{1} - p_{1})^2,\; (l_{1} + p_{2})^2,\\
			l_{2}^2 - m^2,\; (l_{2} + p_{1})^2,\; (l_{2} - p_{2})^2
		\end{array}$ \\
		fam3 & $\begin{array}[t]{@{}l@{}}
			(l_{1} + l_{2} - p_{1} + p_{2})^2,\; l_{1}^2 - m^2,\; (l_{1} - p_{1})^2,\\
			(l_{1} + p_{2})^2,\; l_{2}^2 - m^2,\; (l_{2} + p_{2})^2,\; (l_{2} - p_{1})^2
		\end{array}$ \\
		fam4 & $\begin{array}[t]{@{}l@{}}
			(l_{2} + p_{1})^2-m^2,\; l_{2}^2,\; (l_{1} - l_{2} - p_{1} - p_{2})^2-m^2,\\
			(l_{1} - l_{2} - p_{1})^2,\; l_{1}^2 - m^2,\; (l_{1} - p_{2})^2,\; (l_{1} - p_{1})^2
		\end{array}$ \\
		fam5 & $\begin{array}[t]{@{}l@{}}
			(l_{1} + l_{2} - p_{1} - p_{2})^2,\; l_{1}^2 - m^2,\; (l_{1} - p_{1})^2,\\
			(l_{1} - p_{2})^2,\; l_{2}^2 - m^2,\; (l_{2} - p_{2})^2,\; (l_{2} - p_{1})^2
		\end{array}$ \\
		fam6 & $\begin{array}[t]{@{}l@{}}
			l_{1}^2 - m^2,\; l_{2}^2,\; (l_{2} - p_{1})^2,\;
			(l_{1} - p_{2})^2,\\ (l_{1} - l_{2} - p_{2})^2 ,\; (l_{1} - l_{2} + p_{1} - p_{2})^2,\; 
			(l_{1} - l_{2})^2
		\end{array}$ \\
		\hline
	\end{tabular}
	\caption{The five generic integral families appearing in the 
	forward Compton scattering amplitudes. 
		Families~4 and~5 are obtained from families~1 and~3, respectively, by the crossing $p_2 
		\to 
		-p_2$.}
	\label{tab:integralFamilies}
\end{table}

We organize the scalar integrals into integral families, each defined by a 
linearly independent set of propagators. To map a given integral onto one of these families 
efficiently, we employ the method of Ref.~\cite{Pak:2011xt} and implement its algorithms into a 
computer code, which identifies equivalent propagator sets through momentum shifts. In formulation of \cite{Pak:2011xt}, 
a momentum shift corresponds to a permutation of the 
Feynman parameters in the underlying Feynman graph polynomial. Our code systematically searches 
for such permutations, mapping all scalar integrals onto the six families listed in 
Tab.~\ref{tab:integralFamilies}. Finally, we reduce the integrals belonging to each family to a 
set of master integrals using integration-by-parts (IBPs) identities generated by 
\texttt{REDUZE}~\cite{vonManteuffel:2012np}, obtaining 111 generic master integrals in total. 
The amplitude then takes the form
\begin{equation}\label{eq:formFactorsStandardForm}
	\mathcal{M}_{l}(Q^2,\, m^2,\, s;d) = \sum_{k} r_{lk}(Q^2,\, 
	m^2,\, s;d) \, j_k,
\end{equation}
where $l$ labels the form factor (cf. Eq.~\eqref{eq:amplTensorStr}), $r_{lk}$ are rational 
functions of the kinematic variables and $d$, and $j_k$ denote the master integrals.

Equation~\eqref{eq:formFactorsStandardForm} decomposes the amplitude into rational coefficient
functions and master integrals, rendering its analytic structure transparent. The rational
coefficients are straightforward, except for the Born-cut diagrams, where the imaginary part is
defined by cutting only the massless propagator $1/(s+\mathrm{i}0)$, which appears entangled 
with IBP coefficients. To access the relevant analytic properties, one must consider the real 
and imaginary parts of the generic master integrals $j_k$, as discussed in
Refs.~\cite{Grigo:2013rya,Grigo:2014oqa,Hoschele:2012xc,Hoschele:2013pvt}. In this work, we do 
not compute the full generic integrals; instead, we extract specific contributions to their 
real and imaginary parts using cut‑based techniques, as explained below.
\section{Computation of master integrals}\label{sec:diffEqsCompute}
The computation of coefficient functions requires the integration of dimensionally regulated scalar integrals over the 
kinematic domain defined by the phase-space constraints of the final states. We categorize these integrals into four 
distinct sets based on their cut structure: the I- and II-systems (real-real), III-system (mixed real-real and 
real-virtual), and the IV-system (fully virtual). Given the similarity, in the real-radiation case, between differential 
equation systems, we focus on providing full computational details for the I-system, while presenting the II- and 
III-systems mainly to highlight essential differences and to enable reproduction of our results. The IV-system is 
treated separately, though many aspects of the calculations discussed for the real-radiation case apply readily to the 
virtual one.
\subsection{Cut based approach to differential equations}\label{sec:cutApproach}
The differential equation method represents a standard approach for evaluating Feynman 
integrals \cite{Kotikov:1991pm,Bern:1993kr}. To derive the differential equations (DEQs), we 
begin by differentiating the Feynman integrals with respect to each kinematic variable in 
Eq.~\eqref{eq:invariants}. Using IBPs computed earlier, we express the differentiated integrals 
in terms of our chosen basis, obtaining a system of coupled differential equations of the form
\begin{equation}\label{eq:genericDiffEqs}
	\pdv{\boldsymbol{j}}{x_i} = M_i \cdot \boldsymbol{j},
\end{equation}
where the kinematic variables have been rendered dimensionless: $x_1=s/Q^2,x_2=m^2/Q^2$. The 
vector $\boldsymbol{j}=(j_1,j_2,\ldots,j_{111})$ represents the \textit{generic} master integrals in 
the chosen basis.  Here, the matrices $M_i=M_i(x_1,x_2;\eps)$ for $i=1,2$ contain rational functions 
of the dimensionless kinematic variables and the dimensional regularization parameter $\eps$. These matrices 
encode all analytic properties of the integrals, including physical thresholds, singular structures, 
and branch cuts. Crucially, they also fully characterize the \textit{cuts} of the integrals. 

The imaginary part of a generic integral is given by the cutting equations 
\cite{Veltman:1994wz,Hoff:2015kub}
\begin{equation}\label{eq:cuttingEquations}
	\operatorname{Im} \, j_k = j_k - (j_k)^* \doteq
	\theta(x_1 - 4x_2) \, j_{k,\mathrm{RRB}} +
	\theta(x_1 - x_2) \, j_{k,\mathrm{RRC}} +
	\theta(x_1) \, (j_{k,\mathrm{RRA}} + j_{k,\mathrm{RV}}),
\end{equation}
where we use the labels of the corresponding cut Feynman diagrams (RRB, RRC, RRA, RV from 
Tab.~\ref{tab:intermidStates}) as indices for the cut master integrals \(j_{k,i}\) to emphasize 
their correspondence; these distinct mathematical objects should not be confused with each 
other. Here, \((j_k)^*\) denotes the complex conjugate integral. The $\theta$-functions encode the 
physical thresholds for particle production. The cut integrals in 
Eq.~\eqref{eq:cuttingEquations} are obtained by setting the relevant propagators on-shell via 
the replacement
\begin{equation}\label{eq:onShellProps}
	\frac{1}{D_i}=\frac{1}{L_i^2 - M_i^2 + \mathrm{i}0}\to
	\delta^{+}(L_i^2 - M_i^2)=\delta(L_i^2 - M_i^2)\,\theta(L_i^0),
\end{equation}
where $\delta(x)$ is the Dirac delta function.

To identify which propagators must be put on-shell for a given cut, the graph‑theoretic 
approach outlined in the previous section is employed. For each master integral, the 
corresponding graph is constructed, its weighted adjacency matrix is formed, and it is 
processed with the same algorithms. The results are presented in Tab.~\ref{tab:cutRules}, which 
lists the sets of propagator indices (from the families in Tab.~\ref{tab:integralFamilies}) 
that must be placed on-shell for each cut.

We now apply the technical ingredients presented above to organize the calculation:
\begin{enumerate}
	\item For the $\mathrm{I}$-cut ($\mathrm{I}\equiv\mathrm{RRB}$), we enforce the 
	corresponding on-shell conditions from Eq.~\eqref{eq:onShellProps} within the generic 
	system of Eq.~\eqref{eq:genericDiffEqs}, using the mapping provided in 
	Tab.~\ref{tab:cutRules}. This produces a reduced differential system
	\begin{equation}\label{eq:IcutDEQs}
		\pdv{\boldsymbol{j}_{\mathrm{I}}}{x_i} = M_i^{\mathrm{I}} \cdot \boldsymbol{j}_{\mathrm{I}},
	\end{equation}
	where $\boldsymbol{j}_{\mathrm{I}}=(j_{\mathrm{I},1},\ldots,j_{\mathrm{I},27})$ are the master 
	integrals for the $\mathrm{I}$-cut. The matrices $M_i^{\mathrm{I}}$ are obtained by removing from 
	$M_i$ the rows and columns corresponding to integrals that do not admit this specific cut.
	
	\item Similar to the previous case, we discard generic $j$ that do not possess the 
	$\mathrm{RRC}$ cut. Consequently, we obtain the second reduced differential system
	\begin{equation}\label{eq:IIcutDEQs}
		\pdv{\boldsymbol{j}_{\mathrm{II}}}{x_i} = M_i^{\mathrm{II}} \cdot 
		\boldsymbol{j}_{\mathrm{II}},
	\end{equation}
	where $\boldsymbol{j}_{\mathrm{II}}=(j_{\mathrm{II},1},\ldots,j_{\mathrm{II},8})$ are the master 
	integrals for the $\mathrm{II}$-cut. The matrices $M_i^{\mathrm{II}}$ are obtained in the same 
	manner as in the case of $\mathrm{I}$-system.
	
	\item For the $\mathrm{III}$-system, we consider contributions in the kinematic region 
	defined by $0 < x_1 < x_2$, forbidding the $\mathrm{I}$ and $\mathrm{II}$ 
	cuts. The relevant contributions arise from the union of the $\mathrm{RRA}$ and 
	$\mathrm{RV}$ 
	cuts, i.e., $\mathrm{III} \equiv \mathrm{RRA} \cup \mathrm{RV}$. We therefore select the 
	subset of generic master integrals whose imaginary part receives contributions from these cuts 
	within this kinematic domain. This leads to the system
	\begin{equation}\label{eq:IIIcutDEQs}
		\pdv{\boldsymbol{j}_{\mathrm{III}}}{x_i} = M_i^{\mathrm{III}} \cdot 
		\boldsymbol{j}_{\mathrm{III}},
	\end{equation}
	where $\boldsymbol{j}_{\mathrm{III}}=(j_{\mathrm{III},1},\ldots,j_{\mathrm{III},32})$ 
	comprises the relevant generic master integrals. The system is defined implicitly by the 
	kinematic restriction, rather than by enforcing a specific on-shell condition.
\end{enumerate}
Integrating the differential equations~(\ref{eq:IcutDEQs}, \ref{eq:IIcutDEQs}, \ref{eq:IIIcutDEQs}) 
gives the phase-space master integrals required for the real-radiation contributions to the 
coefficient functions in Eq.~\eqref{eq:hadronStructureFunctions} (e.g. \textrm{RRA},\text{RRB}, 
\text{RRC}, and \textrm{RV}).

The integrals corresponding to the virtual cut in Tab.~\ref{tab:intermidStates} are also 
derived from the generic differential system of Eq.~\eqref{eq:genericDiffEqs}. However, the 
cut-based method used for the real-radiation case above cannot be applied directly. In Feynman diagrams 
with the \textrm{VV} cut, the massless quark propagator $1/(x_1 + \mathrm{i}0)$ mixes with 
coefficients from the IBP reduction, preventing a straightforward application of the on-shell 
prescription. Consequently, at the level of scalar integrals, no direct virtual cut exists. 
Instead, we identify the virtual contribution by selecting Feynman diagrams with the virtual 
cut \textrm{VV}, collecting the master integrals appearing in their amplitudes, and discarding 
the others, obtaining:
\begin{equation}\label{eq:IVcutDEQs}
	\pdv{\boldsymbol{j}_{\mathrm{IV}}}{x_i} = M_i^{\mathrm{IV}} \cdot \boldsymbol{j}_{\mathrm{IV}},
\end{equation}
where $M_i^{\mathrm{IV}}$ is a $41 \times 41$ matrix derived from the generic one in 
Eq.~\eqref{eq:genericDiffEqs}, and the vector $\boldsymbol{j}_{\mathrm{IV}}$ contains the 
generic integrals needed for the virtual contribution. These integrals are then further 
processed: as we demonstrate later, it suffices to study the real part of the 
$\mathrm{IV}$-system in the asymptotic limit $s \to 0$ and extract the leading term of the 
corresponding expansion.

Finally, the canonical form of differential equations introduced by Henn~\cite{Henn:2013pwa} provides 
significant computational advantages. For a given integral family $k$, the canonical 
differential equation reads
\begin{equation}
	\pdv{\boldsymbol{J}_{k}}{y_i} = \eps\, S^{k}_{i}\,\boldsymbol{J}_{k},
\end{equation}
and can be obtained through a gauge transformation
\begin{equation}
	\begin{aligned}
		\boldsymbol{j}_{k} &= T_{k}\,\boldsymbol{J}_{k},\\
		S^{k}_{i} &= T_k^{-1}\cdot M^{k}_i\cdot T_{k} - T_k^{-1}\cdot\partial_{y_i}T_k.
	\end{aligned}
\end{equation}
Here, $\boldsymbol{j}_{k}$ denotes the vector of Laporta master integrals discussed previously, while 
$\boldsymbol{J}_{k}$ represents the vector of canonical master 
integrals for $k=\mathrm{I, II, III}$. The transformation matrices $T_k$ convert the original 
$\eps$-dependent matrices $M^{k}_i=M^{k}_{i}(y_1,y_2;\eps)$ into canonical matrices 
$S^{k}_{i}=S^{k}_{i}(y_1,y_2)$ that are independent of the dimensional regularization 
parameter. The new kinematic variables $y_1$ and $y_2$ are introduced later.

The canonical form offers two key advantages. First, the general solution can be expressed 
compactly as a path-ordered exponential
\begin{equation}\label{eq:canonGeneralSol}
	\boldsymbol{J}_{k} = \mathrm{P}\exp\left(\epsilon\int_{\gamma}\sum_{k=1,2}
	S_k \cdot \mathrm{d} y_k\right),
\end{equation}
where $\mathrm{P}$ is the path-ordering operator and $\gamma$ specifies the 
integration contour. Second, it allows a straightforward expansion in the $\epsilon\to0$ 
limit, for instance, via Picard iteration. Thus, the problem of solving the differential 
equations reduces to finding the transformations $T$ that bring them into this canonical form.
\subsection{Real-radiation: canonical transformations} 
\label{sec:transformationEpsForm}
The construction of appropriate transformations that bring the given system of DEQs to 
canonical form presents considerable challenges and requires a careful analysis of the 
intricate singular structure of the $M_i^k$ matrices. To address this issue, we employ the 
framework implemented in the \texttt{Mathematica} package \texttt{Libra}~\cite{Lee:2020zfb}, 
which provides semi-automatic tools for obtaining canonical forms. The package implements 
algorithms based on balance transformations and the systematic reduction of the Poincar\'{e}
rank of the DEQs, ultimately ensuring that all singularities are 
Fuchsian~\cite{Lee:2014ioa}.

A notable feature of \texttt{Libra} is the possibility, within limitations, to 
find balance transformations involving algebraic extensions
\begin{equation*}
	T_{i,j}=R_1(x)+u(x)R_2(x),
\end{equation*}
where $R_{1,2}$ are rational functions and $u$ satisfies $u^2=q(x)$, where $q(x)$ is a polynomial. Such 
extensions in the transformation matrix are often related to massive particles in the scattering process, as heavy-pair 
production.

Transformations of the system \eqref{eq:genericDiffEqs} would have involved three square 
roots, defined by:
\begin{equation}\label{eq:myAlgExtensions}
	\begin{aligned}
		u_1^2 - (x_1^2+2x_1(1-x_2)+(1+x_2)^2)&=0,\\
		(x_1 +2(1+x_2) )u_2^2 - x_1+2(1-x_2)&=0,\\
		(x_1-4x_2)u_3^2 -x_1&=0.
	\end{aligned}
\end{equation}
By imposing on-shellness and other constraints as previously detailed, $u_2$ is completely 
removed from our systems, while the algebraic extensions (roots) $u_{1}$ and $u_{3}$ remain and 
still pose significant computational challenges.

To manage the remaining roots, we implement the variable transformation 
\begin{equation}\label{eq:variableChange}
	x_1\to\frac{y_1}{(1-y_1)(1-y_2)},\,x_2\to\frac{y_2}{(1-y_1)(1-y_2)},
\end{equation}
which rationalizes $u_1$, leaving only $u_3$ in the system $M^{I}$. The resulting differential 
equations in the $y$-variables read
\begin{equation}\label{eq:newVarsDiffSystem}
	\pdv{\boldsymbol{j}_{k}}{y_i}=M_i^{k}\cdot\boldsymbol{j}_{k},
\end{equation}
where now $M_i^{k}=M_i^{k}(y_1,y_2),\, k=1,2,3$ and 
$\boldsymbol{j}_{k}=\boldsymbol{j}_{k}(y_1,y_2)$. 

This variable change is chosen to achieve three goals: parametrizing the soft 
limit $x_1\to0$ in a convenient form (discussed later), not increasing the 
complexity of the coefficient matrices, and reducing the algebraic complexity of the transformations. Consequently, 
after applying this change, no further technical 
obstacles prevent us from obtaining the $\eps$-form of Eq.~\eqref{eq:newVarsDiffSystem}. The canonical differential 
equation matrices in Fuchsian form read\footnote{Given that 
our system involves algebraic extensions, we consider singularities as Fuchsian provided no 
subsequent variable transformation increases their Poincaré rank. We thank M.~Bonetti for 
private conversations on this matter.}
\begin{equation}\label{eq:fuchsForm}
	\begin{aligned}
		S_k^{\mathrm{I}}(y_1,y_2)&=\sum_i\frac{S^{\mathrm{I}}_{i k}}{y_k-a^{\mathrm{I}}_{k,i}}+ 
		u(y_1,y_2)\sum_i\frac{S^Q_{i k}}{y_k-a_{k,i}^{Q}},\\
		S_k^{\mathrm{II}}(y_1,y_2)&=\sum_i\frac{S^{\mathrm{II}}_{i 
				k}}{y_k-a^{\mathrm{II}}_{k,i}},\\
		S_k^{\mathrm{III}}(y_1,y_2)&=\sum_i\frac{S^{\mathrm{III}}_{i 
		k}}{y_k-a^{\mathrm{III}}_{k,i}},
	\end{aligned}
\end{equation}
where the $S_k^{l}$ are constant matrices, $u_3\equiv u = \sqrt{y_1/(y_1-4y_2)}$, and the 
$a_i^{l}$ denote the locations of singularities in kinematic space, often referred to as 
\textit{letters} in the literature. A set of letters constitutes an \textit{alphabet} of the 
corresponding DEQ system. The relevant alphabets are presented in 
Eqs.~\eqref{eq:ab_1}--\eqref{eq:ab_4}.

\subsection{Real-radiation: integration}\label{sec:integrationPSI}
As discussed previously, the canonical form of the differential equations admits an exponential 
solution with $\eps$ appearing in a fully factorized form. We expand the exponential solution 
in Eq.~\eqref{eq:canonGeneralSol} as a series around $\eps=0$, using the canonical matrices 
from Eq.~\eqref{eq:fuchsForm}. The practical implementation of this expansion employs Picard 
iteration.\footnote{Picard iteration reproduces the formal solution of 
Eq.~\eqref{eq:canonGeneralSol} when summed to all orders in $\eps$, provided the matrix is in 
block-triangular form.} The solution for the $i$-system on the path $\gamma_1^{i} : 
[0,y_{k_i}]$ is obtained by successive iterations of the corresponding matrix $S_{k_i}^{i}$, 
where $i = \mathrm{I},\mathrm{II},\mathrm{III}$ and 
$\{k_{\mathrm{I}}=2,\;k_{\mathrm{II}}=2,\;\;k_{\mathrm{II}}=1\}$, as 
follows:
\begin{equation}\label{eq:solutionPicard}
	\boldsymbol{J}_{i} = \sum_{n=0}^{N_{i}} \eps^n \boldsymbol{J}_{i}^{(n)},
\end{equation}
with
\begin{equation}\label{eq:picardIterate}
	\boldsymbol{J}^{(n)}_{i} = 
	\tilde{\boldsymbol{C}}^{(n)}_{\gamma_1^{i}} + 
	\int_{0}^{y_{k_i}} S^{i}_{k_i}(t) \circ 
	\boldsymbol{J}^{(n-1)}_{i}(t)\,\mathrm{d}t,
\end{equation}
where $\circ$ denotes path-ordered matrix multiplication along $\gamma_1^{i}$. We employ a 
compact notation for the $S$-matrices and other tensors, showing only the active integration 
variable (i.e., $t \equiv (y_1, t)$ for $i = \mathrm{I},\mathrm{II}$ and $t \equiv (t, y_2)$ for $i = 
\mathrm{III}$, with the inactive coordinate held fixed). Here, for $n=0,1,2,\ldots$, 
$\tilde{\boldsymbol{C}}^{(n)}_{\gamma_1^{i}}$ are integration constants and 
$\boldsymbol{J}^{(0)}_{i} \equiv \tilde{\boldsymbol{C}}^{(0)}_{\gamma_1^{i}}$. It is sufficient to expand up to order  
$N_{i}=5$ to capture all finite contributions in the coefficient functions, since the highest pole of typical 
four-point NNLO corrections does not exceed $\mathcal{O}(\eps^{-4})$~\cite{Catani:1998bh}.

The iterated integral structure of the evolution matrix is made explicit by re-expanding the 
Picard iterates. For the straight-line path \(\gamma_1^{i}\) we have
\begin{equation}\label{eq:evolutionStep1}
	\boldsymbol{J}_{i} = \operatorname{U}_{\gamma_1^{i}} \cdot 
	\boldsymbol{C}_{\gamma_1^{i}},
\end{equation}
where the evolution matrix \(\operatorname{U}_{\gamma_1^{i}}\) is given by the series
\begin{equation}\label{eq:picardIter}
	\begin{aligned}
		\operatorname{U}_{\gamma_1^{i}} &= \mathds{1} + \sum_{n\geq 1} \eps^n S^{i}_{(n)}(y_1,y_2), 
		\\
		S^{i}_{(n)}(y_1,y_2) &= \int_{0}^{y_{k_i}} \mathrm{d}t_n \int_{0}^{t_n} 
		\mathrm{d}t_{n-1} 
		\cdots \int_{0}^{t_2} \mathrm{d}t_1 \,
		S_{k_i}^{i}(t_n) \circ S_{k_i}^{i}(t_{n-1}) \circ \cdots \circ S_{k_i}^{i}(t_1),
	\end{aligned}
\end{equation}
with \(\mathds{1}\) the identity matrix and the matrices in the integrand are path-ordered 
along \(\gamma_1^{i}\).

\begin{figure}[!tbp]
	\centering
	\includegraphics[width=.5\textwidth]{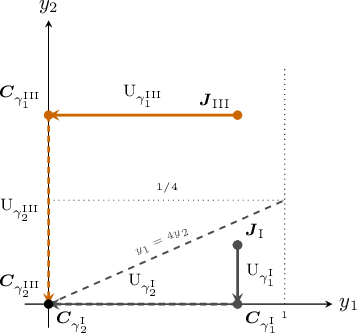}
	\caption{Integration paths in the $(y_1,y_2)$ plane specific to the systems defined in the 
	text. The $\mathrm{I}$- and $\mathrm{II}$-systems follow the same integration contour, 
	differing only in their initial phase-space points.}
	\label{fig:integrationPath}
\end{figure}

We demonstrate the method for determining integration constants using the $\mathrm{I}$-system 
as an example \cite{Lee:2020obg}. To obtain 
$\boldsymbol{C}_{\gamma_1^{\mathrm{I}}}=\boldsymbol{C}_{\gamma_1^{\mathrm{I}}}(y_1)$, we 
construct the corresponding differential system in two steps.

First, because the matrices $S^{\mathrm{I}}_{2}$ are Fuchsian, the evolution matrix 
$\operatorname{U}_{\gamma_1^{\mathrm{I}}}$ from Eq.~\eqref{eq:evolutionStep1} admits near 
$y_2=0$ a generalized series expansion, whose coefficients are determined via the Frobenius 
method \cite{barkatou2010frobenius, Lee:2017qql, Lee:2020zfb}. For the $\mathrm{I}$-system, 
this expansion takes the form
\begin{equation}\label{eq:asymptoticSol}
	\operatorname{U}_{\gamma_1^{\mathrm{I}}}(y_1,y_2) = \sum_{\lambda\in S} 
	(y_{2})^{\lambda}\,\sum_{n=0}^{\infty}\,\sum_{k=0}^{K_{\lambda}}A_{n+\lambda,k}\,y_2^n 
	\,\log^{k}(y_2),
\end{equation}
where $A_{n+\lambda,k}=A_{n+\lambda,k}(y_{1},\eps)$ are rational function matrices. The 
exponents $\lambda$ and their multiplicities $K_{\lambda}$ are obtained from the corresponding 
indicial polynomial as prescribed in Ref.~\cite{barkatou2010frobenius}.

Second, substituting this series expansion into the canonical $y_2$-system (with the DEQ 
	matrices provided in Eq.~\eqref{eq:fuchsForm}) gives
\begin{equation}\label{eq:boundaryEqs}
	\pdv{\boldsymbol{C}_{\gamma_1^{\mathrm{I}}}}{y_1} = \eps\, \tilde{S}^{\mathrm{I}}(y_1)
	\cdot\boldsymbol{C}_{\gamma_1^{\mathrm{I}}} ,\quad 
	\tilde{S}^{\mathrm{I}}(y_1)=\sum_i\frac{\tilde{S}^{\mathrm{I}}_i}{y_1-b^{\mathrm{I}}_i}.
\end{equation}
Here, $\tilde{S}^{\mathrm{I}}$ is a new matrix derived from the canonical ones in 
Eq.~\eqref{eq:fuchsForm}, with its alphabet defined in Eq.~\eqref{eq:ab_4}. Provided the 
original system is in canonical form, this two-step procedure preserves the $\eps$-form of the 
differential equations. Consequently, they can be solved using the same methods applied to the 
$y_2$-system.

Integrating Eq.~\eqref{eq:boundaryEqs} along the $\gamma_{2}^{\mathrm{I}}$ path gives
\begin{equation}\label{eq:picardIter2}
	\begin{aligned}
		\boldsymbol{C}_{\gamma_1^{\mathrm{I}}} &= 
		\operatorname{U}_{\gamma_2^{\mathrm{I}}}^{\mathrm{I}} 
		\cdot \boldsymbol{C}_{\gamma_2^{\mathrm{I}}} = \left( \mathds{1} + \sum_{n \geq 1} 
		\eps^n 
		\tilde{S}^{\mathrm{I}}_n(y_1) \right) \cdot \boldsymbol{C}_{\gamma_2^{\mathrm{I}}}, \\
		\tilde{S}^{\mathrm{I}}_n(y_1) &= \int_{0}^{y_{1}} \mathrm{d}t_n \int_{0}^{t_n} 
		\mathrm{d}t_{n-1} 
		\cdots \int_{0}^{t_2} \mathrm{d}t_1 \,
		\tilde{S}^{\mathrm{I}}(t_n) \circ \tilde{S}^{\mathrm{I}}(t_{n-1}) \circ \cdots \circ 
		\tilde{S}^{\mathrm{I}}(t_1),
	\end{aligned}
\end{equation}
where $\circ$ now denotes path-ordering along $\gamma_2^{\mathrm{I}}$. The path 
$\gamma_{2}^{\mathrm{I}} : [0, y_1]$ connects the solution at $\boldsymbol{y} = (y_1, 0)$ from 
Eq.~\eqref{eq:picardIter} to the origin, where the integration constants 
$\boldsymbol{C}_{\gamma_2^{\mathrm{I}}}$ are fixed via the PSLQ method (discussed later). 
Applying the same procedure in the order specific to the $\mathrm{II}$, $\mathrm{III}$ systems 
produces the derived matrices $\tilde{S}^{\mathrm{II}}$ and $\tilde{S}^{\mathrm{III}}$ with alphabet 
defined in Eqs.~(\ref{eq:ab_6},~\ref{eq:ab_7}), respectively; the corresponding differential 
equations 
are then integrated analogously to Eq.~\eqref{eq:picardIter2}.

The matrix elements in Eqs.~\eqref{eq:picardIter} and \eqref{eq:picardIter2} are iterated integrals 
defined as 
\cite{chen1977iterated}:
\begin{equation}\label{eq:goncharovIterIntDef}
	G(a_1,\ldots,a_n;y_k) := \int_{0}^{y_{k}} \frac{\mathrm{d}^*t_1}{t_1-a_1}
	\int_{0}^{t_1} \frac{\mathrm{d}^*t_2}{t_{2}-a_{2}}
	\cdots \int_{0}^{t_{n-1}} \frac{\mathrm{d}^*t_n}{t_n - a_n},
\end{equation}
with $G(;y_k) := 1$. We adopt the notation of Ref.~\cite{Lee:2021iid}, where the integration 
measure $\mathrm{d}^*t$ is generalized to include 
both rational and algebraic cases:
\begin{equation}
	\mathrm{d}^* t = \begin{cases}
		\mathrm{d}t & \text{(rational)} \\
		u\,\mathrm{d}t & \text{(algebraic)}
	\end{cases}, \qquad u=\sqrt{y_1/(y_1 - 4 y_2)}.
\end{equation}
When all measures are rational, Eq.~\eqref{eq:goncharovIterIntDef} coincides with the standard 
definition of Goncharov polylogarithms \cite{goncharov2001multiple}. The algebraic case, which 
appears exclusively in $\operatorname{U}_{\gamma_1^{\mathrm{I}}}$, requires additional 
treatment. We apply the transformation
\begin{equation}\label{eq:squareRoot}
	t_i \to y_1\frac{(u_i^2-1)}{4u_i^2},
\end{equation}
to rationalize the measure. After simplifying the resulting integrand, the expression again 
reduces to Goncharov polylogarithms, but with a modified alphabet given in Eq.~\eqref{eq:ab_6}.

\subsection{Virtual: canonical transformations and integration}\label{sec:virtualIntegrals}
Virtual integrals require special treatment because we cannot isolate virtual cuts at the scalar 
integral level, as discussed in Sec.~\ref{sec:cutApproach}. The complication arises because the 
would-be-cut massless quark propagator becomes entangled with IBP coefficients, 
preventing direct extraction of the desired cut.
 
Our systematic solution retains the original scalar propagator structure by exploiting how virtual 
contributions originate from the Feynman propagator's analytic structure. Specifically, the 
imaginary 
part of the amplitude (Eq.~\eqref{eq:formFactorsStandardForm}) stems, in the virtual case, from 
rational coefficients through the on-shell condition defined in Eq.~\eqref{eq:onShellProps}
\begin{equation}\label{eq:onshellStrangeQuark}
	\frac{1}{y_1+\mathrm{i}0} \to -\mathrm{i}\pi 
	\delta(y_1).
\end{equation} 
Consequently, only the real part of $\boldsymbol{j}_{\mathrm{IV}}$ near $y_1=0$ must be 
computed. For clarity, we apply this substitution only when the pole $1/y_1$ becomes manifest 
in the amplitude, after virtual integrals are substituted in 
Eq.~\eqref{eq:formFactorsStandardForm}. 

To achieve this, we first regularize $M_i^{\mathrm{IV}}$ from Eq.~\eqref{eq:newVarsDiffSystem}
at $y_1 = 0$ by applying the balance transformations obtained using \texttt{Libra}, which
reduce the $1/y_1$ pole to first order. As in the previous section, generalized series
solutions can be obtained via the Frobenius method near simple poles. Eventually, for the
Laporta master integrals, we obtain the solution near $y_1=0$
\begin{equation}
	\boldsymbol{j}_{\mathrm{IV}} = \operatorname{U}_{\gamma_1^{\mathrm{IV}}}
	\cdot \boldsymbol{c}_{\gamma_1^{\mathrm{IV}}},
\end{equation}
where the path $\gamma^{\mathrm{IV}}_1$ matches that of the $\mathrm{III}$-system. The vector
$\boldsymbol{c}_{\gamma_1^{\mathrm{IV}}} = \boldsymbol{c}_{\gamma_1^{\mathrm{IV}}}(y_2)$
contains the integration constants for the generic integrals near $y_1 = 0$. For the virtual
contribution, we retain only the \(\lambda = 0\) solution from the full expansion (cf.
Eq.~\eqref{eq:asymptoticSol}), since the other solutions do not contribute to the virtual
integrals. Thus, the expansion reduces to the form
\begin{equation}\label{eq:asymptoticSolVirt}
	\operatorname{U}_{\gamma_1^{\mathrm{IV}}}(y_1, y_2) =
	\sum_{n = 0}^{N} A_{n, k} \, y_1^n + \mathcal{O}(y_1^{N+1}),
\end{equation}
which we truncate at order \(N\). This truncation retains only the simple-pole contribution in
\(y_1\) at the amplitude level; we have verified explicitly that higher-order poles cancel.
Terms with $\lambda \neq 0$ would lead to integrands proportional to
\begin{equation}
	\delta(y_1)\,(y_1)^{\lambda} \log^{k}(y_1),
\end{equation}
which, upon imposing the $\delta$-constraint, vanish identically, provided that $\lambda\propto\operatorname{Re}(\eps) < 
0$. Here,
we obtain 16 independent integration constants in
$\boldsymbol{c}_{\gamma_1^{\mathrm{IV}}}$, while the remaining 26 boundary conditions of the original
system are discarded, as they do not contribute to the virtual part.

Similar to the real-radiation case, we derive differential equations for the virtual 
integration constants by substituting Eq.~\eqref{eq:asymptoticSolVirt} into the corresponding 
$y_2$-equations from 
Eq.~\eqref{eq:newVarsDiffSystem}. This produces a simple single-variable system, that is 
transformed to $\eps$-form using the procedure detailed in Sec.~\ref{sec:transformationEpsForm}
\begin{equation}\label{eq:boundaryEqsVirt}
	\pdv{\boldsymbol{C}_{\gamma_1^{\mathrm{IV}}}}{y_2} = 
	\eps\, \tilde{S}^{\mathrm{IV}}(y_2) \cdot 
	\boldsymbol{C}_{\gamma_1^{\mathrm{IV}}}, \quad 
	\tilde{S}^{\mathrm{IV}}(y_2) = 
	\sum_i \frac{\tilde{S}^{\mathrm{IV}}_i}{y_2 - b^{\mathrm{IV}}_i}.
\end{equation}
The notation is kept consistent with the real-radiation case. The virtual system's alphabet is 
given in Eq.~\eqref{eq:ab_V}.

Integration proceeds analogously to the real-radiation case using the Picard iteration method 
from Sec.~\ref{sec:integrationPSI} (Eq.~\eqref{eq:picardIter2}).

\subsection{Boundary conditions}\label{sec:boundaryCond}
The differential equations for the $\mathrm{I}$-, $\mathrm{II}$-, $\mathrm{III}$-, and 
$\mathrm{IV}$-systems have been integrated analytically up to integration constants. 
This section describes the numerical determination of these constants. These numerically 
determined constants are subsequently employed in Sec.~\ref{sec:integrationPSI} to obtain the 
fully normalized, uniformly transcendental expressions for the master integrals

We determine the constants using the PSLQ algorithm~\cite{Ferguson:1999}, which identifies 
rational linear combinations of basis constants matching high-precision numerical values. The 
relevant basis consists of constants of transcendental weight up to five~\cite{Henn:2013pwa}:
\begin{equation}\label{eq:PSLQbasis}
	\pi^{n},\, \zeta_n,\, \log^n(2),\, \mathrm{Li}_n(1/2),\qquad n = 1, 2, 3, 4, 5,
\end{equation}
and their products, where weight adds under multiplication. The weight corresponds to the depth of 
iterated integration in the functions' definition (cf. Eq.~\eqref{eq:goncharovIterIntDef}). The 
constants above admit known definitions in terms of iterated integrals from which their 
weights can be determined.

Our final expressions are required to be in \emph{uniformly transcendental} (UT) form: all terms 
at order $\eps^n$ have weight exactly $n$. The canonical basis alone does not guarantee this which 
necessitates an appropriate $\eps$-dependent normalization. We determine the normalization factors 
$f_i$ by analyzing the simplest integral topology within each system. We transform the simplest 
Laporta master integrals, computed via parametric integration, to the canonical basis. 
Extracting their boundary conditions from the generalized series solutions 
(Sec.~\ref{sec:integrationPSI}) produces constants. Demanding that the resulting constants 
be of pure weight determines the normalization factor in front of the integral measure
\begin{equation}\label{eq:normalizationTotal}
	f_i\,S_{\eps}^2\,\int 
	\frac{\mathrm{d}^{d}l_1}{(2\pi)^{d'}}\frac{\mathrm{d}^dl_2}{(2\pi)^{d'}},
\end{equation}
with $d' = d-1$ for the $\mathrm{I}$, $\mathrm{II}$  systems (cut integrals) and $d' = d$ otherwise. 
The factor $f_i$ for each system is given in Eq.~\eqref{eq:normalizationFactor} of the appendix. The 
spherical factor $S_{\epsilon}$ is defined as
\begin{equation}
	S_{\epsilon} = \exp\left(\epsilon(\ln4\pi-\gamma_{\mathrm{E}})\right),
\end{equation}
where $\gamma_{\mathrm{E}} \approx 0.577\ldots$ is the Euler–Mascheroni constant.

High-precision numerical values for the Laporta integrals 
$\boldsymbol{j}$ are obtained using \texttt{AMFlow}~\cite{Liu:2022chg}, which implements the 
auxiliary mass flow method and supports arbitrary-precision arithmetic (typically 
$\mathcal{O}(300)$ digits). The kinematics are chosen according to the cut prescription discussed 
in Sec.~\ref{sec:cutApproach}.
\begin{itemize}
	\item $\mathrm{I}$-system: $y_1 > 4y_2,\; y_2 > 0$ (denominators put on-shell).
	\item $\mathrm{II}$-system: $y_1 > y_2,\; y_2 > 0$ (denominators put on-shell).
	\item $\mathrm{III}$-system: $y_1 < y_2,\; y_2 > 0$ (generic integrals; keep imaginary part).
	\item $\mathrm{IV}$-system (virtual): $y_1 = 0,\; y_2 > 0$ (boundary from 
	Eq.~\eqref{eq:asymptoticSolVirt}).
\end{itemize}

The constants are extracted via the following procedure:
\begin{enumerate}
	\item Compute the Laporta integrals $\boldsymbol{j}$ numerically at the chosen kinematic 
	points.
	\item Apply the inverse balance transformation  together with the normalization factors of 
	Eq.~\eqref{eq:normalizationTotal} to convert the 
	numerical $\boldsymbol{j}$ to the normalized UT basis $\boldsymbol{J}$.
	\item Evaluate the analytic solutions for $\boldsymbol{J}$ (Sec.~\ref{sec:integrationPSI}) to 
	the 
	same high precision using \texttt{PolyLogTools}~\cite{Duhr:2019tlz} (which calls 
	\texttt{GiNaC} \cite{Bauer:2000cp}).
	\item Subtract the analytic result from the numerical one. All functional dependence 
	cancels, leaving only constants.
	\item Use the PSLQ algorithm to fit each derived numerical value to a linear combination of 
	transcendental constants presented in Eq.~\eqref{eq:PSLQbasis}.
\end{enumerate}
We conclude this section by outlining how our results were verified. 
First, we note that the fitting procedure already guarantees the correctness of the evolution 
part. Indeed, if the analytic solution were incorrect, the resulting numerical constants would 
not match the transcendental basis defined in Eq.~\eqref{eq:PSLQbasis}, and the PSLQ algorithm 
would fail to produce a fit.

As a further check, we evaluated our final analytic solutions at kinematic points different from those used in the 
fitting step, and compared them against corresponding numerical results from
\texttt{AMFlow}, finding full agreement within specified accuracy reaching $300$ digits.
\section{Results}\label{sec:results}
Here we present the main result of this paper: the renormalized coefficient functions through 
$\mathcal{O}(\alpha_s^2)$. First, we briefly discuss UV renormalization  and the regularization 
of IR singularities. We then detail the exact structure of the partonic structure functions at 
each perturbative order. Complete analytic expressions are provided as auxiliary files with 
this publication.

\subsection{Renormalization}\label{sec:renorm}
The forward Compton scattering amplitude from Eq.~\eqref{eq:opticalTheorem2} admits a 
perturbative expansion in the bare strong coupling constant $\alpha_{s,0}$
\begin{equation}
	\mathcal{M}_{0}^{\mu\nu} = 
	\mathcal{M}_{0}^{(0),\mu\nu}+
	a_0\,
	\mathcal{M}_{0}^{(1),\mu\nu}+
	a_0^2\,
	\mathcal{M}_{0}^{(2),\mu\nu}
	+ \mathcal{O}(a_0^3),
\end{equation}
where $a_0 = \alpha_{s,0}/2\pi$. The terms $\mathcal{M}_{0}^{(k),\mu\nu}$ with $k=0,1,2$ denote 
the respective perturbative contributions; the leftmost subscript “$0$” indicates an 
unrenormalized quantity. This expansion matches, order by order in $a_0$, the perturbative 
structure of the unrenormalized coefficient functions that enter 
Eq.~\eqref{eq:partonicPerturbExp}. Below we summarize the renormalization constants needed to 
obtain finite quantities.

The choice of renormalization scheme for the strong coupling constant requires careful 
consideration. While one could, in principle, use the $\overline{\mathrm{MS}}$ scheme, the 
resulting DGLAP evolution becomes rather involved due to heavy-quark self-energy diagrams entering 
the corresponding equations. It is preferable to subtract the heavy-quark contribution at zero 
momentum \cite{Collins:1978wz}. We therefore adopt the decoupling scheme, following 
Refs.~\cite{Qian:1984kf, Buza:1995ie, Bojak:2000eu, Bierenbaum:2007qe}: here, the light flavors 
are treated in $\overline{\mathrm{MS}}$, while the heavy quark decouples from the evolution 
equations.

Within this scheme, the coupling renormalization is given by
\begin{equation}\label{eq:alpha_renorm}
		a = Z_g^2 a_0 = a(Q^2)
		\left(
			1 + a(Q^2)
			\left(
				-\frac{\beta_0}{2\eps}
				-n_H\frac{\beta_{0,Q}}{2\eps}%
				\left(
					\frac{m^2}{Q^2}%
				\right)^{-\eps}%
				\left(
					1+\frac{1}{2}\eps^2\frac{\pi^2}{6}
				\right)
			\right)
		 \right),
\end{equation}
where the lowest-order beta functions are
\begin{align}
	\beta_0 &= \frac{11}{3}C_A - \frac{4}{3}T_R n_L, \label{eq:3NFScheme}\\
	\beta_{0,Q} &= -\frac{4}{3}T_R .
\end{align}
Here, $n_L=3$ and $n_H=1$ denote the number of light and heavy quarks, respectively, and the 
$\mathfrak{su}(N)$ color factors are $C_F=(N_c^2-1)/2N_c,\;C_A=N_c,\;T_R=1/2$ with $N_c=3$. The 
term proportional to $n_H$ modifies the renormalization constant $Z_g$ by introducing a 
mass-dependent finite counterterm. This modifies the $1/\epsilon$ pole structure in the 
Callan-Symanzik equation, resulting in a beta function from which the heavy-quark contribution is 
removed. Using this scheme ensures that only the $n_L$ light flavors participate in the DGLAP 
evolution.

The renormalized amplitude reads
\begin{equation}\label{eq:renormComptonAmplitude}
	\mathcal{M}^{\mu\nu} = 
	\left.\left(Z^{\mathrm{OS}}_{2,Q}
	\mathcal{M}_{0}^{\mu\nu}\right)\middle|_{\begin{array}{@{}c@{\,}c@{}l@{}}
			\scriptstyle a_0 & \scriptstyle \to & \scriptstyle Z_g^2 a, \\[-3pt]
			\scriptstyle m_0 & \scriptstyle \to & \scriptstyle Z_m^{\mathrm{OS}} m .
	\end{array}}
	\right.
	= \mathcal{M}^{(0),\mu\nu} + a\,\mathcal{M}^{(1),\mu\nu} + a^2\,\mathcal{M}^{(2),\mu\nu}
	+ \mathcal{O}(a^3),
\end{equation}
where $Z^{\mathrm{OS}}_{2,Q}$ and $Z^{\mathrm{OS}}_m$ are the on-shell renormalization 
constants 
taken from Ref.~\cite{Melnikov:2000qh}. We have verified that at this perturbative order the 
choice of $Z_{2,Q}$ does not affect the coefficient functions provided the pole mass $m$ is 
used. We believe that this is due to ``the $S$-matrix being the same for renormalized as for unrenormalized 
Green functions'', as follows from the discussion in Ref.~\cite{Sterman:1993hfp}. The 
renormalization scheme for the operator matrix elements that define PDFs must, however, be 
consistent with our choice.

The finite flavor-stripped coefficient functions inherit the perturbative expansion of the 
renormalized 
amplitude
\begin{equation}\label{eq:renormCoefFun}
			C_{j,n}(y_1,y_2,Q^2)= \delta(y_1) + a\,C^{(1)}_{j,n}(y_1,y_2,Q^2) + 
			a^2\,C^{(2)}_{j,n}(y_1,y_2,Q^2) + 
			\mathcal{O}(a^3),
\end{equation}
where
\begin{equation}
	C^{(i)}_{j,n} = 
	\frac{\mathcal{N}_{j}}{\mathcal{B}_{j}}\operatorname{Im}\mathcal{M}^{(i)}_{j,n}.
\end{equation}
Here, the normalization factors $\mathcal{N}_{j}$ are defined in Eq.~\eqref{eq:normOmega} and 
$\mathcal{B}_{j}$ are the Born coefficient functions, presented later. This normalization 
choice implies $C^{(0)}_{j,n}\equiv \delta(y_1)$. The imaginary parts 
$\operatorname{Im}\mathcal{M}^{(i)}_{j,n}$ are extracted using the cut‑based methods described 
in Secs.~(\ref{sec:computationDetails},~\ref{sec:cutApproach}). 

Finally, we discuss the impact of the renormalization scheme on the coupling, with coefficient 
functions obtained at the factorization/renormalization point $\mu^2_F=\mu_R^2=Q^2$, 
i.e., $C^{(i)}_{j,n}\equiv C^{(i)}_{j,n}(\mu_R^2=Q^2)$. To restore their full scale dependence, 
we use the fact that the hadronic structure functions in 
Eq.~\eqref{eq:hadronStructureFunctions} are independent of the renormalization scale. Within 
our scheme, the scale dependence is straightforward to derive; we present the result without 
reproducing the full evolution equations
\begin{multline}
	C_{j,n}= \delta(y_1) + a(\mu_R^2)\,C^{(1)}_{j,n}(Q^2)\\ + 
	a^2(\mu_R^2)\Bigl(C^{(2)}_{j,n}(Q^2) + \beta_0 
	\log\!\Bigl(\frac{\mu_F^2}{Q^2}\Bigr)\,C^{(1)}_{j,n}(Q^2)\Bigr) + 
	\mathcal{O}(a^3),
\end{multline}
where, to restore the dependence on the factorization scale, we have solved the 
renormalization group equation for $\alpha_s$ to express $\alpha_s(\mu_F^2)$ in terms of 
$\alpha_s(\mu_R^2)$. Here, $\beta_0$ (Eq.~\eqref{eq:3NFScheme}) corresponds to $n_L$ light 
flavors. 
\subsection{Isolating soft contribution}\label{sec:softReg}
In Sec.~\ref{sec:computationDetails}, we used dimensional regularization to regulate both UV and 
IR divergences in the forward Compton scattering amplitude up to $\mathcal{O}(\alpha_s^2)$. At 
this order, the corresponding loop integrals produce poles up to $1/\epsilon^4$ at two 
loops~\cite{Catani:1998bh}. While UV divergences are removed by renormalization in the previous section, the 
cancellation of IR poles, in general, requires a systematic analysis of phase-space integrals, as different kinematic 
configurations produce characteristic $\eps$-pole structures. Here, we only need to focus on the soft limit $y_1 \to 0$ 
of the coefficient functions.

Dimensional regularization correctly handles soft-collinear 
divergences in the $y_1\to0$ limit. However, our treatment of the 
$\epsilon\to0$ limit during master integral evaluation disturbs 
their proper behavior near $y_1=0$, which in turn affects the 
partonic structure functions in the soft region. This becomes 
evident from the asymptotic expansion of the master integrals 
$\boldsymbol{J}_{\mathrm{III}}$ in the soft limit
\begin{equation}\label{eq:correctAsym}
	\operatorname{A}_{y_1}\left[\boldsymbol{J}_{\mathrm{III}}\right] 
	= \sum_{\lambda\in S} 
	y_{1}^{\lambda}\,\sum_{n=0}^{N}\,\sum_{k=0}^{K_{\lambda}}\,y_1^n 
	\,\log^{k}(y_1)A^{\mathrm{III}}_{n+\lambda,k}\cdot%
	\boldsymbol{C}_{\gamma_1^{\mathrm{III}}},
\end{equation} 
with exponent set 
$S=\{-\eps,-2\eps,-3\eps,-4\eps\}$. Here, 
$\operatorname{A}_{y_1}$ denotes the truncated generalized series 
solution from Sec.~\ref{sec:integrationPSI}. This expansion makes 
the integrals' singular structure explicit, ensuring that physical 
soft singularities are properly regulated once inserted into the 
hadronic integral (cf. Eq.~\eqref{eq:hadronStructureFunctions}).

The appearance of $y_1^{-a-b\eps}$ divergences for $a,\,b>0$, which enter 
the hadronic structure functions through Eq.\eqref{eq:correctAsym}, signals the 
non-commutativity of the limits $y_1 \to 0$ and $\eps \to 0$, 
requiring a careful ordering. Following Ref.~\cite{Hoff:2015kub}, we 
restore the correct soft behavior via the regularization prescription
\begin{equation}\label{eq:regulatePS}
	\boldsymbol{J}_{\mathrm{III}} \to \boldsymbol{J}_{\mathrm{III}} 
	- \operatorname{T}_{y_1}\left[\boldsymbol{J}_{\mathrm{III}}\right] 
	+ \operatorname{A}_{y_1}\left[\boldsymbol{J}_{\mathrm{III}}\right],
\end{equation}
for the $\mathrm{III}$-system. Here, $\operatorname{T}_{y_1}$ is the 
truncated Taylor expansion around $y_1 = 0$. The subtraction of the 
Taylor expansion removes the incorrect behavior of the scalar 
integrals in the soft limit, while adding back the asymptotic 
expansion reinstates the correct singular structure. The precise 
truncation order $N$ is unimportant in practice, provided it is 
large enough not to affect finite contributions in the coefficient 
functions.

Applying the prescription in Eq.~\eqref{eq:regulatePS} regularizes 
the corresponding structure functions through the replacement rule
\begin{equation}\label{eq:refIRMassless}
	\lim_{\Delta\to 0}\int_{\Delta}^{1}\dd y_1\, g\,y_1^{-a-b\eps} = 
	\int_{0}^{1}\dd y_1 g\Biggr( \frac{\delta(y_1)}{1-a-b\eps}+y_1^{1-a}
	\sum_{k=0}\frac{(-b\,\eps)^k}{k!}\mathcal{D}_k(y_1)\Biggl),
\end{equation}
where $\Delta$ is an auxiliary infrared regulator, introduced to properly define the integral 
on the left-hand side; it completely vanishes as $\Delta\to 0$, provided that 
$\operatorname{Re}(\eps)<0$.  Here, $g = g(y_1)$ is some properly defined test function. The 
functions $\mathcal{D}_k(x)$ are \emph{plus-distributions} defined as~\cite{Altarelli:1979ub}:
\begin{equation}\label{eq:plusDistr}
	\int_{0}^{1}\dd x\, g(x)\mathcal{D}_k(x) = \int_{0}^{1} \dd x\, 
	\frac{\log^k(x)}{x}(g(x)-g(0)),
\end{equation}
for $k\geq 0$; for $k=-1$ we additionally define 
$\mathcal{D}_{-1}(x)=\delta(x)$. Through 
Eq.~\eqref{eq:refIRMassless}, the soft divergences of the amplitude 
are re-expressed as poles in $\epsilon$, as required.\footnote{To 
avoid proliferating definitions, we use the same symbol for lower 
and upper plus-distributions. The position of the regulated 
divergence can be inferred from the variable context (e.g., $\xi$, 
$\xi'$, or $y_1$).}

The $\mathrm{RRB}$ and RRC contributions behave differently in the soft 
region. The phase-space constraint for the heavy quark final 
state(s), imposed via $\theta(s-4m^2)$ or $\theta(s-m^2)$ in the convolution integral 
(Eq.~\eqref{eq:ansatzFactorization}), shields it from the massless 
soft singularity at $s=0$. The heavy-quark mass thus acts as a 
natural IR cutoff, allowing a naive $\eps$-expansion of the 
$\mathrm{I}$ and II systems.
\begin{figure}[!tbp]
	\centering
	\includegraphics[width=0.44\textwidth]{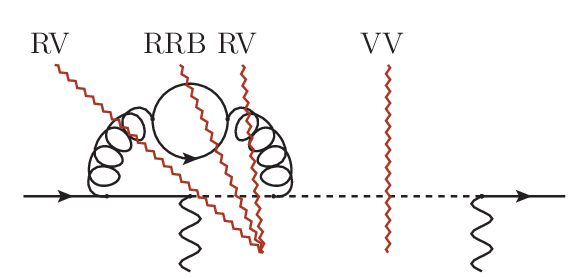}
	\caption{One of the diagrams in which each individual cut (cf. Tab.~\ref{tab:intermidStates}) produces a large 
	triple logarithm.}
	\label{fig:SudakovDiagram}
\end{figure}

Although beyond the scope of this work, it is instructive to 
consider the limit~$\tfrac{m^2}{Q^2} \to 0$. Convolving PDFs with the 
$\mathrm{RRB}$ coefficient functions produces hadronic functions that 
diverge as $\log^3(m^2/Q^2)$ in this limit due to Sudakov enhancement 
\cite{Rijken:1995gi,Buza:1997mg,Forte:2010ta}. One diagram 
responsible for this behavior is shown in 
Fig.~\ref{fig:SudakovDiagram}. 
The explanation follows Refs.~\cite{Buza:1997mg,Forte:2010ta}: as 
$m^2/Q^2$ decreases, the upper integration boundary in 
Eq.~\eqref{eq:ansatzFactorization} approaches the massless soft 
singularity at $\xi=1$, and the virtual gluon producing the heavy-quark pair becomes soft, inducing an additional 
logarithmic 
enhancement. However, the sum over all cut diagrams of the type represented in 
Fig.~\ref{fig:SudakovDiagram}, must cancel these logarithms exactly (cf. Refs.~\cite{Rijken:1995gi, 
Buza:1997mg, Forte:2010ta}).

To facilitate an analytical demonstration of this cancellation later, we prepare by slightly 
modifying the regularization procedure for the $\mathrm{I}$-integrals in the soft limit as
\begin{equation}\label{eq:regulatePSMod}
	\boldsymbol{J}_{\mathrm{I}} \to \boldsymbol{J}_{\mathrm{I}} 
	- \operatorname{T}_{y_1}\circ\operatorname{A}_{y_2}\left[\boldsymbol{J}_{\mathrm{I}}\right] 
	+ \operatorname{A}_{y_1}\circ\operatorname{A}_{y_2}\left[\boldsymbol{J}_{\mathrm{I}}\right].
\end{equation}
We emphasize that, due to phase-space constraints, the limits $y_1\to0$ and $y_2\to0$ do not commute
for the $\mathrm{I}$-system. Applying this prescription to the $\mathrm{RRB}$ diagrams gives an
asymptotic expansion of the corresponding coefficient functions of the form
\begin{equation}
	g = g_0 + g_1\;y_2^{-\eps}+ g_2\;y_2^{-2\eps} + \mathcal{O}(y_2),
\end{equation}
where $g_l = g_l(y_1)$ for $l=0,1,2$. The coefficients obtained in this way are correctly regulated
in the soft $y_1\to0$ limit. The convolution integral for each coefficient can then be written 
as
\begin{multline}\label{eq:plusDistrMod}
	\int_{4y_2}^{1}\dd y_1\,g_l(y_1)\,y_1^{-1-b\eps}
	= \\ \int_{0}^{1}\dd y_1\, g_l(y_1) 
	\Biggl(
	\delta(y_1)\frac{1-4^{-b\eps}y_2^{-b\eps}}{-b\eps}
	+ \sum_{k=0}^{\infty}\frac{(-b\,\eps)^k}{k!}\mathcal{D}_k(y_1)
	\Biggr)
	+ \mathcal{O}(y_2),
\end{multline}
where the lower limit in the second term has been set to $0$.

A few comments on this result are necessary. In the previously considered case of massless 
final states (the III-system), an auxiliary infrared regulator $\Delta$ is introduced to 
properly define 
convolution integrals; this auxiliary regulator vanishes in the $\Delta \to 
0$ limit for $\operatorname{Re}(\eps)<0$. For the $\mathrm{RRB}$ cut, however, the lower 
boundary in Eq.~\eqref{eq:plusDistrMod} is fixed by phase-space constraints and represents an 
\emph{actual}, not auxiliary, regulator. Consequently, it must be retained in the subsequent 
analysis of the RRB contribution. The non-commutativity of infrared 
limits~\cite{Gottschalk:1980rv} means that applying the auxiliary-regulator limiting procedure 
here would incorrectly suppress the $y_1^{-a-b\eps}$ contribution as $y_2 \to 0$, generating a 
spurious $1/\eps$ pole in the integrand. The correct 
procedure is therefore to expand in $\eps$ \emph{under the integral sign} while keeping $y_2$ 
finite but vanishingly small; in this way, the $\eps$-pole cancels, and the logarithmically 
enhanced terms required for a smooth massless limit are produced~\cite{Gottschalk:1980rv}. 
\subsection{Born coefficient functions}\label{sec:bornRes}
Upon applying the variable change defined in Eq.~\eqref{eq:variableChange}, the on-shell
condition (cf. Eq.~\eqref{eq:onshellStrangeQuark}) reduces to
\begin{equation}
	\delta(s) = \frac{(1-y_2)}{Q^2}\delta(y_1),
\end{equation}
which, together with Eqs.~\eqref{eq:strfun_partonic_def} and \eqref{eq:normOmega}, provides the
leading order hadron structure functions
\begin{equation}\label{eq:bornStrFun}
	\begin{aligned}
		F_1^0(x_B,Q^2) &= |\mathcal{V}_{\mathrm{cs}}|^2 \mathcal{B}_{1} f_Q(\eta,Q^2) = 
		\tfrac{1}{2}S_{+} |\mathcal{V}_{\mathrm{cs}}|^2 f_Q(\eta,Q^2), \\
		F_2^0(x_B,Q^2) &= |\mathcal{V}_{\mathrm{cs}}|^2 \mathcal{B}_{2} f_Q(\eta,Q^2) = 
		x_BS_{+}/(1-y_2) |\mathcal{V}_{\mathrm{cs}}|^2 f_Q(\eta,Q^2), 
		\\
		F_3^0(x_B,Q^2) &= |\mathcal{V}_{\mathrm{cs}}|^2 \mathcal{B}_{3} f_Q(\eta,Q^2) = 
		2R_{+} |\mathcal{V}_{\mathrm{cs}}|^2 f_Q(\eta,Q^2).
	\end{aligned}
\end{equation}
Here, we follow the notation of Ref.~\cite{Kretzer:1998ju}, where $R_+ = S_+/2 = g^2/16$. The 
coefficients $\mathcal{B}_{i}$ in Eq.~\eqref{eq:bornStrFun} are the Born coefficient functions 
$C_{i}^{(0)}$ that enter the normalization condition Eq.~\eqref{eq:renormCoefFun}.
\subsection[$\mathcal{O}(\alpha_s)$ coefficient functions]{\texorpdfstring{$\bm{\mathcal{O}(\alpha_s)}$}{} coefficient 
functions}\label{sec:resultsNLO}
We did not discuss the calculation of the $\mathcal{O}(\alpha_s)$ corrections, but they are 
relatively straightforward to obtain following the same steps as in the higher-order 
calculation detailed in this paper.

The coefficients are given for the flavor-stripped tensors defined in
Eq.~\eqref{eq:flavorDecompostion}. At this order, the only nonvanishing
contributions are proportional to $|V_{\mathrm{cs}}|^2$ (and $|V_{\mathrm{cd}}|^2$); all other 
CKM structures vanish. The corresponding CKM
factors are reinstated when required. The structure of the
$\mathcal{O}(\alpha_s)$ contribution to the coefficient functions is
\begin{equation}\label{eq:NLOStrFun}
	C_{j,n}^{(1)}%
	= C_F\biggl(c_{j,n}^{(1),f}
	+ \sum\limits_{k={-1,1}}^{1}\mathcal{D}_{k}(y_1)\,c_{j,n}^{(1),k}\biggr).
\end{equation}
where $C_{j,\bar{\mathrm{Q}}}^{(1)} = 0$ and the $n={\mathrm{Q}}$ coefficients read
\begin{equation}
	\begin{aligned}
		c_{j,Q}^{(1),1} &= -2,\\[4pt]
		c_{j,Q}^{(1),0} &= -\frac{7 + 4\,\G(0,y_2)}{2}, \\[4pt]
		c_{j,Q}^{(1),-1} &=S_{j,Q}^{(1)}+V_{j,Q}^{(1)} = -2\,\G(1,y_2)\,\G(0,y_2)  - 
		\frac{3}{2}\,\G(0,y_2) \nonumber \\
		&\quad	+ 2\,\G(0,1,y_2) - \frac{\pi^2}{3} - \frac{5}{2} + 
		(1-\delta_{2j})\,\frac{y_2}{1-y_2}\,\G(0,y_2),
	\end{aligned}
\end{equation}
for $j=1,2,3$. The first two coefficients are in fact universal for all $j$; the Kronecker 
delta selects the term for $j\neq2$. The remaining \textit{finite} coefficients 
$c_{j,n}^{(1),f}$ take the following form
\begin{align}
	c_{1,Q}^{(1),f} &= \frac{2\G(0,y_1y_2)\, A_1 - (1-y_1 y_2) B_1}{2(1 - y_1y_2 )^3}, \\
	c_{2,Q}^{(1),f} &= \frac{2\G(0,y_1y_2)\, A_2 - (1-y_1 y_2) B_2}{2(1 - y_1)(1 - y_1y_2 
	)^3},\\
	c_{3,Q}^{(1),f} &= \frac{2\G(0,y_1y_2)\, A_3 - (1-y_1 y_2) B_3}{2(1 - y_1y_2 )^2},
\end{align}
with
\begin{align*}
	A_1 &= (1 + y_1 y_2)\, A_3 + 2 \left(1-2 y_2\right) y_2 y_1^2+4 y_2 \left(4 y_2-1\right)y_1, 
	\\[4pt]%
	A_2 &= y_1 (y_2 - 1)\, A_1 + \left(14 y_2^2-17 y_2+4\right)\, y_2 y_1^3 \\%
	& -2\left(13y_2^2-17 y_2+8\right)\, y_2 y_1^2 + \left(-2 y_2^2 + 16 y_2-1\right)\, y_1 - %
	8 y_2 + 2, 
	\\[4pt]%
	A_3 &= \left(-2 y_2^2+2 y_2-1\right)\,  y_2 y_1^2 + \left(6y_2-1\right)\, y_1 - 6 y_2+2,
\end{align*}
and
\begin{align*}
	B_1 &=\left( y_1 y_2 - \frac{23}{7} \right)\, B_3 + \frac{2}{7} \left(53 
	y_1-150\right)-\frac{2}{7} 
	\left(7 y_1^2 + 94 y_1 - 128\right)\, y_2, \\%
	B_2 &= \left( y_1 y_2 + \frac{24 - 11y_1}{7} \right)\, B_1 + \frac{4}{7} \left(4 y_1^2-59 
	y_1+126\right)\, y_2^2y_1\\
	& -\frac{2}{7} \left(10 y_1^2-143 y_1+127\right)\, y_2 y_1 - \frac{2}{7} \left(10 y_1^2+9 
	y_1-71\right),\\%
	B_3 &= 9 y_2 + \left(7 y_2^2-10 y_2+4\right)\, y_1  - 10.
\end{align*} 

It is straightforward to verify that our $\mathcal{O}(\alpha_S)$ result reduces to 
the $\overline{\mathrm{MS}}$ one in the limit $m^2/Q^2\to0$. 
Following Ref.~\cite{Kretzer:1998ju} and using their variables that are better suited to study 
the high-virtuality/low-mass limit (Eq.~\eqref{eq:varsKretzer}), we obtain in the 
aforementioned limit
\begin{multline}\label{eq:NLOasymptoticRes}
	\lim_{m \to 0} \int_{\eta}^1\frac{\dd\xi'}{\xi'}f_{Q}\left(\frac{\eta}{\xi'},\mu_F^2\right)
	C_{j,Q}^{(1)}\left(\xi',\,\frac{m^2}{Q^2},\,\mu_R^2\right)
	=\\
	\int_{\eta}^1\frac{\dd\xi'}{\xi'}f_{Q}\left(\frac{\eta}{\xi'},\mu^2_F\right)
	\Biggl[
	C_{j,Q}^{\overline{\mathrm{MS}},(1)}\left(\xi',\,\frac{Q^2}{\mu_F^2},\,\mu_R^2\right)
	+
	\alpha_S(\mu_R)C_{j,Q}^{\overline{\mathrm{MS}},(0)} \,
	\Gamma^{(0)}_{QQ}(\xi',\mu_F,m)
	\Biggr].
\end{multline}
We have verified that our $C_{j,Q}^{\overline{\mathrm{MS}},(i)}$ functions are the same as, 
e.g., the ones presented in Ref.~\cite{Moch:1999eb} (and references therein). The expression 
$\Gamma^{(0)}_{QQ}$ is nothing but the generalized quasi-collinear counter 
term that describes a ``heavy quark inside of a heavy quark'' within ACOT scheme 
\cite{Aivazis:1993kh,Aivazis:1993pi} and corresponds to the so-called PDF matching 
coefficient within the FONLL scheme \cite{Forte:2010ta}
\begin{multline}\label{eq:counterTermACOT}
	\Gamma^{(0)}_{QQ}(\xi',\mu_F,m) = 
	C_F\,\delta(1-\xi')\left(2+\frac{3}{2}\log\frac{\mu_F^2}{m^2}\right)
	+\\
	C_F\,
	\left[\frac{1+\xi'^2}{1-\xi'}\left(\log\frac{\mu_F^2}{m^2(1-\xi')^2} - 1 
	\right) 
	\right]_{+}.
\end{multline}
Unlike Ref.~\cite{Kretzer:1998ju}, we additionally 
accounted for the fermion number conservation at the leading-logarithmic level. Namely, we have 
verified that
\begin{equation*}
	\int_0^1\dd \xi'\,\Gamma^{(0)}_{QQ}(\xi',\mu,m) =0,
\end{equation*}
which, subsequently, ensures that we indeed reproduce the $\overline{\mathrm{MS}}$ result in 
Eq.~\eqref{eq:NLOasymptoticRes}.
\subsection[$\mathcal{O}(\alpha_s^2)$ coefficient functions]{\texorpdfstring{$\bm{\mathcal{O}(\alpha_s^2)}$}{} 
coefficient 
	functions}\label{sec:resNNLO}
At this order, coefficient functions proportional to $|V_{\mathrm{ud}}|^2$ and
$|V_{\mathrm{us}}|^2$ appear for the first time. We therefore restore the CKM matrix
elements and reintegrate them into the coefficient functions, which are no longer
flavor-stripped.

The complexity of the $\mathcal{O}(\alpha_s^2)$ corrections necessitates the decomposition
\begin{equation}\label{eq:NNLOStrFun}
	C_{j,n}^{(2)} = |V_{\mathrm{cd}}|^2 (L_{j,n} + \theta(y_1-4y_2) \, H_{j,n}) + 
	|V_{\mathrm{ud}}|^2\theta(y_1-y_2) \, K_{j,n}, \quad 
	j=1,2,3,
\end{equation}
where the $\theta$ functions implement the heavy-quark production thresholds at
$y_1 = y_2$ and $y_1 = 4y_2$. The light term $L_{j,n}$ receives contributions from the 
$\mathrm{VV}$, $\mathrm{RV}$, and $\mathrm{RRA}$ cuts (see Tab.~\ref{tab:intermidStates}),
the heavy term $H_{j,n}$ receives contributions solely from the $\mathrm{RRB}$ cut,
and the remaining heavy term $K_{j,n}$ arises from the $\mathrm{RRC}$ cut.\footnote{The 
designations ``light'' and ``heavy'' distinguish contributions with light or heavy final 
states, respectively. Both kind of functions retain the full heavy-quark mass dependence.}
For the cuts contributing to $L_{j,n}$, the antiquark-initiated process vanishes
identically, i.e.\ $L_{j,\bar{Q}}=0$.

The color structure of the light coefficient is
\begin{equation}
	L_{j,\mathrm{Q}} = C_F^2 \, L_{1,j} + C_A C_F \, L_{2,j} + n_L C_F \, L_{3,j} + n_H C_F \, 
	L_{4,j},
\end{equation}
where each color-stripped function $L_{i,j}$ ($i=1,\ldots,4$) admits an expansion in 
plus-distributions analogous to the $\mathcal{O}(\alpha_s)$ case,
\begin{equation}
	L_{i,j} = c^{(2),f}_{i,j,L}
	+ \sum_{k=-1}^{3} \mathcal{D}_{k}(y_1) \, c^{(2),k}_{i,j,L},
\end{equation}
with the distributions $\mathcal{D}_k$ defined as before. In the following, we present only the 
plus-distribution coefficients $c_{i,j,L}^{(2),k}$ for $k=-1,0,\ldots,3$ while the soft 
structure of $H/\bar{H}$ is discussed later. Note that the $n_H C_F$ term, while associated 
with heavy-quark loops, contributes to $L_{j,\mathrm{Q}}$ via the $\mathrm{RV}$ cut of diagrams 
with a heavy-quark loop insertion (see Fig.~\ref{fig:SudakovDiagram}).

The plus-distribution coefficients for $k=1,2,3$ read
\begin{equation}
	\begin{aligned}
		c_{1,j,L}^{(2),3} &= 2, \\ 
		c_{1,j,L}^{(2),2} &= \frac{21 + 12\,\G(0,y_2)}{2}, \\
		c_{2,j,L}^{(2),2} &= \frac{11}{2}, \\
		c_{3,j,L}^{(2),2} &= -1, \\[4pt]
		c_{1,j,L}^{(2),1} &= 4 \G(0,y_2)^2
		+ \bigl(4 \G(1,y_2) + \tfrac{15 y_2-17}{y_2-1}\bigr)\G(0,y_2)
		- 4 \G(0,1,y_2) \\
		&\quad + \frac{69}{4}
		+ \delta_{2j}\,\frac{2y_2}{1-y_2}\,\G(0,y_2), \\[4pt]
		c_{2,j,L}^{(2),1} &= \frac{22}{3} \G(0,y_2) - \frac{11}{3} \G(1,y_2)
		+ 2\zeta(2) + \frac{95}{36}, \\[4pt]
		c_{3,j,L}^{(2),1} &= -\frac{4}{3} \G(0,y_2) + \frac{2}{3} \G(1,y_2)
		- \frac{13}{18}.
	\end{aligned}
\end{equation}
For $k=0$ the coefficients are
\begin{equation}
	\begin{aligned}
		c_{1,j,L}^{(2),0} &=
		\bigl(4 \G(1,y_2) + \tfrac{y_2-3}{y_2-1}\bigr) \G(0,y_2)^2 \\
		&\quad + \bigl(7 \G(1,y_2) - 4 \G(0,1,y_2)
		+ \tfrac{41-27 y_2}{4-4 y_2}\bigr) \G(0,y_2)
		- 7 \G(0,1,y_2) \\
		&\quad - 2\zeta(3) + 3\zeta(2) + \frac{67}{8}
		+ \delta_{2j}\,\frac{y_2}{1-y_2}\,
		\frac{\G(0,y_2)\bigl(4\G(0,y_2)+7\bigr)}{2}, \\[4pt]
		c_{2,j,L}^{(2),0} &=
		\frac{11}{6} \G(0,y_2)^2 - \frac{11}{3} \G(1,y_2)\G(0,y_2)
		+ 2\zeta(2)\G(0,y_2) \\
		&\quad - \frac{34}{9} \G(0,y_2) - \frac{77}{12} \G(1,y_2)
		+ \zeta(3) + \frac{17}{3}\zeta(2) - \frac{905}{72}, \\[4pt]
		c_{3,j,L}^{(2),0} &=
		-\frac{1}{3} \G(0,y_2)^2 + \frac{2}{3} \G(1,y_2)\G(0,y_2)
		+ \frac{4}{9} \G(0,y_2) \\
		&\quad + \frac{7}{6} \G(1,y_2) - \frac{2}{3}\zeta(2) + \frac{85}{36}.
	\end{aligned}
\end{equation}
Here, most coefficients are independent of $j$ with the exception: the $C_F^2$ contribution
($i=1$) to $C_{2,L}^{(2)}$ contains an additional term proportional to
$\delta_{2j}$ (Kronecker delta) which is similar to the previous order.

The $k=-1$ (delta‑function) coefficients, together with the regular finite parts
$c_{i,j,L}^{(2),f}$, are too lengthy to present here and are provided in the auxiliary
material described later. The leading coefficient of the Taylor expansions of the delta‑term 
coefficients in the $m^2/Q^2\to 0$ limit, however, are compact
\begin{equation}\label{eq:lightCoefDelta}
\begin{aligned}
	c_{1,j,L}^{(2),-1} &= \left(\frac{9}{8} - 2\zeta(2)\right) 
	\log^2\left(\frac{m^2}{Q^2}\right)
	+ \left(-\zeta(2) - 2\zeta(3) + \frac{27}{8}\right) \log\left(\frac{m^2}{Q^2}\right) \\
	&\quad + \frac{1}{8}\bigl(110\zeta(2) - 54\zeta(4) - 88\zeta(3) + 95\bigr)
	- 12\zeta(2)\log(2)+ \mathcal{O}\left(\frac{m^2}{Q^2}\right), \\
	c_{2,j,L}^{(2),-1} &= \frac{11}{8} \log^2\left(\frac{m^2}{Q^2}\right)
	+ \left(-\frac{11}{3}\zeta(2) + 3\zeta(3) - \frac{35}{8}\right) 
	\log\left(\frac{m^2}{Q^2}\right) \\
	&\quad + \frac{1}{72}\bigl(-1610\zeta(2) + 819\zeta(4) + 804\zeta(3) - 1081\bigr)\\
	&\quad + 6\zeta(2)\log(2) + \mathcal{O}\left(\frac{m^2}{Q^2}\right), \\
	c_{3,j,L}^{(2),-1} &=  -\frac{1}{4} \log^2\left(\frac{m^2}{Q^2}\right)
	+ \left(\frac{2}{3}\zeta(2) + \frac{3}{4}\right) \log\left(\frac{m^2}{Q^2}\right) \\
	&\quad + \frac{1}{36}\bigl(82\zeta(2) - 24\zeta(3) + 71\bigr) + 
	\mathcal{O}\left(\frac{m^2}{Q^2}\right), \\
	c_{4,j,L}^{(2),-1} &= \frac{1}{9} \log^3\left(\frac{m^2}{Q^2}\right)
	+ \frac{19}{18} \log^2\left(\frac{m^2}{Q^2}\right)
	+ \left(\frac{2}{3}\zeta(2) + \frac{265}{54}\right) \log\left(\frac{m^2}{Q^2}\right) \\
	&\quad + \frac{5}{9}\zeta(2) - \frac{2}{3}\zeta(3) + \frac{7951}{648} + 
	\mathcal{O}\left(\frac{m^2}{Q^2}\right),
\end{aligned}
\end{equation}
where we have again used the variable change from Eq.~\eqref{eq:varsKretzer}. We emphasize the 
term $c_{4,j,L}^{(2),-1}$, proportional to $n_H C_F$, diverges as 
$\log^3\left(\frac{m^2}{Q^2}\right)$ as the ratio grows, as anticipated in 
Sec.~\ref{sec:softReg}. The soft-divergent diagrams were shown prior in Fig.~\ref{fig:SudakovDiagram}. 
The coefficient $L_{4,j}$ contributes only to the $k=-1$ (delta function) term and all its 
plus-distribution as well as finite coefficients vanish: $c_{4,j,L}^{(2),-1} = 0$ for $k\geq 0$.

The heavy coefficients admit the following color decomposition
\begin{align}
	H_{j,\mathrm{Q}} &= C_F\left(C_F - \frac{C_A}{2}\right) H_{1,j} + C_F\bigl(H_{2,j} + n_H 
	H_{3,j}\bigr),\\
	H_{j,\bar{\mathrm{Q}}} &= C_F\left(C_F - \frac{C_A}{2}\right) \bar{H}_{1,j} + 
	C_F\bigl(\bar{H}_{2,j} + n_H \bar{H}_{3,j}\bigr).
\end{align}
Here, the sea contributions are the same, $H_{2,j}=\bar{H}_{2,j}$. As 
discussed in Sec.~\ref{sec:softReg}, for $m \neq 0$, the heavy coefficient functions 
$H/\bar{H}_{i,j}(\xi', m)$ themselves are finite (non-singular) for all $\xi'$ in the physical 
region. Therefore, they do not require a plus-prescription regularization to be integrated.

However, the soft divergence emerges in the $m^2/Q^2 \to 0$ limit. The resulting large triple 
logarithm is generated solely by the term proportional to $n_H C_F$. To isolate this singular 
behavior, we consider the asymptotic expansion of $H_{j,n}$ as $m \to 0$. The correct 
asymptotic form is obtained by applying a combination of Taylor and asymptotic expansion 
operators, $\operatorname{T}$ and $\operatorname{A}$, defined as in Eq.~\eqref{eq:regulatePS}. 
Letting $\operatorname{T}_{m^2}$ and $\operatorname{A}_{\xi'}$ denote expansion in $m^2/Q^2$ 
and asymptotic expansion in $\xi'$, respectively, we have
\begin{multline}\label{eq:heavyCoefCorrected}
	\lim_{m \to 0 } \int_{\eta}^{\xi'_{\mathrm{th}}}\frac{\dd 
	\xi'}{\xi'}f_Q\left(\frac{\eta}{\xi'}\right)\,H_{3,j}(\xi',m) 
	\to\\
	\int_{\eta}^{1}\frac{\dd \xi'}{\xi'}f_Q\left(\frac{\eta}{\xi'}\right)\,\left(H_{3,j}^{(0)} 
	- \operatorname{T}_{\xi'}\circ\operatorname{T}_{m^2}[H_{3,j}] + 
	\operatorname{A}_{\xi'}\operatorname{T}_{m^2}\circ[H_{3,j}]\right)+ 
	\mathcal{O}\left(\frac{m^2}{Q^2}\right).
\end{multline}
The term $\operatorname{T}_{m^2} \circ \operatorname{T}_{\xi'}[H_{3,j}]$ subtracts the Taylor 
expansion of $H_{3,j}$ around $\xi'=0$, which contains an artifact behaving as $1/(1-\xi')$ 
that is not singular at finite mass but would otherwise pollute the asymptotic extraction.  The 
structure of the heavy coefficients reads
\begin{align}\label{eq:heavyCoefAsym}
	\operatorname{A}_{\xi'}\operatorname{T}_{m^2}\circ[H_{3,j}] &= \sum_{k=-1}^{2} 
	\mathcal{D}_{k}(y_1) \, c^{(2),k}_{j,H}, \quad \text{where} \\
	c^{(2),2}_{j,H} &= \frac{1}{3}, \\
	c^{(2),1}_{j,H} &= -\frac{2}{3} \log\!\left(\frac{m^2}{Q^2}\right) - \frac{29}{18}, \\
	c^{(2),0}_{j,H} &= \frac{1}{3} \log^2\!\left(\frac{m^2}{Q^2}\right) + \frac{29}{18} 
	\log\!\left(\frac{m^2}{Q^2}\right) - \frac{2\zeta(2)}{3} + \frac{359}{108}, \\
	c^{(2),-1}_{j,H} &= - \frac{1}{9} \log^3\!\left(\frac{m^2}{Q^2}\right) - \frac{29}{36} 
	\log^2\!\left(\frac{m^2}{Q^2}\right) \nonumber \\
	&\quad + \left(\frac{2\zeta(2)}{3} - \frac{359}{108}\right) 
	\log\!\left(\frac{m^2}{Q^2}\right) \nonumber \\
	&\quad - \frac{8 \log^3(2)}{9} + \frac{29 \log^2(2)}{9} + \left(\frac{4\zeta(2)}{3} - 
	\frac{359}{54}\right) \log(2).
\end{align}
Thus, for $s > 4m^2$, combining the $n_H C_F$ terms from the light and heavy 
contributions (Eqs.~(\ref{eq:lightCoefDelta},~\ref{eq:heavyCoefAsym}), respectively), we obtain
\begin{equation}
	L_{4,j} + H_{3,j} \;\longrightarrow\; n_H C_F \, \mathcal{D}_{-1}(\xi') \left( 
	c_{4,j,L}^{(2),-1} + c^{(2),-1}_{j,H} \right) + \ldots \; \approx \; 
	\mathcal{O}\!\left(\log^2\!\left(\frac{m^2}{Q^2}\right)\right),
\end{equation}
so that the potentially large triple logarithm cancels exactly, in analogy with 
Refs.~\cite{Buza:1995ie,Bierenbaum:2007qe}. The inclusion of 
diagrams (specifically VV and RV cuts in Fig.~\ref{fig:SudakovDiagram}) contributing to these 
large logarithms in the region $s < 4m^2$ is subject to the exact definition of the 
corresponding hadron function as discussed in Ref.~\cite{Forte:2010ta}. Our default calculation 
includes these diagrams. A variant excluding their contribution, along with the relevant 
auxiliary files, can be provided upon request.

The color structure of the contributions proportional to $|V_{\mathrm{ud}}|^2$ (and 
$|V_{\mathrm{us}}|^2$) is straightforward, as it is entirely proportional to $C_F$. Apart from 
this, the functions $K_{j,n}$ do not exhibit any additional structure relevant for the present 
discussion; they are expressible in terms of Goncharov polylogarithms with rational
prefactors.

The remaining finite contributions, such as $c^{(2),f}_{i,j,L}$ and their heavy counterparts, 
are too extensive to be listed here. They are provided in full, together with the singular 
coefficients presented above, in an auxiliary file whose structure is described in the 
following section.
\subsection{Auxiliary files and validation}\label{sec:auxANDvalidation}
The analytic expressions for the coefficient functions presented in this publication can be
obtained by cloning the repository from the system command line
\begin{verbatim}
	git clone https://github.com/Qdashkin/CF_HQI_CC_DIS.git
\end{verbatim}
or by downloading the source directly from the GitHub webpage using the URL above. The
repository contains three main files:
\begin{itemize}
	\item \verb|README.md|: a standard \texttt{README} file with a description of the 
	repository.
	\item \verb|strFun.nb|: a \texttt{Mathematica} notebook containing the analytic results
	and documentation.
	\item \verb|strFun.mx|: a \texttt{Mathematica} storage file containing the notebook's
	pre-evaluated definitions for direct loading.
\end{itemize}
The notebook \verb|strFun.nb| includes:
\begin{itemize}
	\item The coefficient functions at leading order (Eq.~\eqref{eq:bornStrFun}).
	\item The $\mathcal{O}(\alpha_s)$ coefficient functions (Eq.~\eqref{eq:NLOStrFun}).
	\item The $\mathcal{O}(\alpha_s^2)$ coefficient functions (Eq.~\eqref{eq:NNLOStrFun}), with 
	the light ($L_{j,n}$) and heavy ($H_{j,n}$ and $K_{j,n}$) terms provided separately.
	\item The correct heavy-flavor soft term (Eq.~\eqref{eq:heavyCoefCorrected}).
\end{itemize}
It also documents all definitions and explicitly notes the (few) instances where the notation 
departs from that used in the preceding sections.

We discuss now the validation of our results through a series of independent checks 
starting with the leading order: our Born results agree exactly with those of 
Ref.~\cite{Aivazis:1993kh}, providing a first consistency test.
 
At $\mathcal{O}(\alpha_s)$, we have verified our expressions against Ref.~\cite{Kretzer:1998ju} 
by expanding their general result in the limit $m_2 \to 0$ and using the prescription of 
Ref.~\cite{Catani:2000ef} to correctly reconstruct the finite 
term.\footnote{Ref.~\cite{Kretzer:1998ju} considers the most general two-flavor case at 
$\mathcal{O}(\alpha_s)$, which includes an extra mass parameter $m_2$.} Furthermore, by 
expanding 
our coefficient functions in the $m \to 0$ limit and 
incorporating the generalized counter‑term from Eq.~\eqref{eq:counterTermACOT}, we have 
explicitly reproduced the well‑known $\mathcal{O}(\alpha_s)$ $\overline{\mathrm{MS}}$ DIS 
results from Ref.~\cite{Bardeen:1978yd} in Eq.~\eqref{eq:NLOasymptoticRes}.

The most robust check of the $\mathcal{O}(\alpha_s^2)$ corrections presented in this work is 
their asymptotic expansion. For the heavy-quark initiated partonic structure functions, it reads
\begin{equation}\label{eq:coefFunAsym}
	C_{0,j,n}^{(2)} = \tilde{C}_{0,j,n}^{(2),0} 
	+ y_2^{-\epsilon}\,\tilde{C}_{0,j,n}^{(2),1} 
	+ y_2^{-2\epsilon}\,\tilde{C}_{0,j,n}^{(2),2} 
	+ \mathcal{O}(y_2),
\end{equation}
where the leading coefficient $\tilde{C}_{0,j,n}^{(2),0}$ is the unrenormalized (bare) 
contribution in the massless-limit. We have verified analytically that this coefficient 
coincides exactly with the unrenormalized partonic structure functions appearing in the purely 
massless calculations. The agreement is not accidental: it is a consequence of the decoupling 
theorem \cite{Appelquist:1974tg} and its refined analysis given in Ref.~\cite{Chetyrkin:1997un} 
within the dimensional regularization that makes this decoupling manifest at the level of 
scattering amplitudes for finite $\eps$.

To obtain the bare massless coefficient functions of Ref.~\cite{Moch:1999eb}, we independently 
recomputed their $\overline{\mathrm{MS}}$ result using the same approach as in the current 
work, thereby obtaining direct access to the bare functions. This independent calculation 
provides an additional, robust verification of our computational framework. Furthermore, 
expanding Eq.~\eqref{eq:coefFunAsym} in the $\epsilon \to 0$ limit reproduces the Taylor 
expansion of our exact (mass‑dependent) bare coefficient functions $C_{0,j,n}^{(2)}$. This 
consistency, together with the numerical check of the master integrals via 
\texttt{AMFlow}~\cite{Liu:2022chg} as discussed in Sec.~\ref{sec:boundaryCond}, further ensures 
the correctness of our results.

We have also computed the bare virtual function, which was not discussed explicitly in the main 
text. In attempting to verify its $\epsilon$-pole structure against the results of 
Refs.~\cite{Mitov:2006xs,Wang:2023qbf}, we were able to match the $\mathcal{O}(\epsilon^{-2})$ 
poles, but found a discrepancy in the $\mathcal{O}(\epsilon^{-1})$ and finite terms. This 
mismatch could arise from an incorrect application of the formulas in those references to our 
specific kinematics, or it may reflect a genuine difference: the diagrams in the cited works do 
not cover the case of a gluon connecting a massive and a massless quark line at the vertex 
(Fig.~\ref{fig:SudakovDiagram}). Regardless of this partial cross-check, we note that within 
our 
calculation the poles from the virtual and soft contributions cancel, and the fully 
renormalized result is finite. This internal consistency provides strong support for the 
correctness of our virtual function and, by extension, the full renormalized coefficient 
functions.

We note that the $\overline{\mathrm{MS}}$ coefficient functions at $\mathcal{O}(\alpha_s^2)$ 
could, in principle, also be extracted by a procedure analogous to that described in 
Sec.~\ref{sec:resultsNLO}. This would, however, require counter‑terms analogous to 
Eq.~\eqref{eq:counterTermACOT} at one higher order, which are not yet available but can be 
computed using methods such as those in Refs.~\cite{Melnikov:2004bm,Bierenbaum:2007qe}. In a 
forthcoming publication \cite{kkudashkin123:2025toappear}, we will instead extract these very 
counter‑terms from the present calculation itself. Indeed, the 
remaining coefficients $\tilde{C}_{0,j,n}^{(2),1}$ and $\tilde{C}_{0,j,n}^{(2),2}$ correspond  
to operator matrix elements entering the operator product expansion discussed in, for instance, 
Ref.~\cite{Buza:1995ie}.
\section{Summary and outlook}\label{sec:conclusions}
We have presented the computational details of the charged-current deep-inelastic 
scattering coefficient functions through $\mathcal{O}(\alpha_s^2)$, retaining full 
heavy-quark mass dependence. Modern variable-flavor number schemes such as ACOT and FONLL 
require $n_L=3$ active flavors as the initial condition in the evolution equations; 
accordingly, we used the decoupling scheme to renormalize the UV divergences, keeping 
only $n_L$ active flavors. We provided a detailed discussion of the structure of the 
coefficient functions and isolated the soft singularities, thereby enabling the subsequent 
convolution with PDFs. The results are provided as a supplementary \texttt{Mathematica} notebook 
file on GitHub.

Although the existence of intrinsic heavy flavor, such as charm, remains an open question, our 
result provides one of the last essential contributions to inclusive DIS at 
NNLO  and will facilitate its determination. In a more conservative 
scenario, our calculation is indispensable for quantifying the underlying theoretical 
uncertainties in current PDF fits or in the future N\textsuperscript{3}LO.  

We expect that, once the remaining PDF matching coefficients, which are a direct outcome of 
this work, become available \cite{kkudashkin123:2025toappear}, a thorough phenomenological 
analysis including the presented results of charged-current DIS will be possible. In 
particular, it will be interesting to probe CC DIS near threshold and at high values of 
virtuality, where the effects of the heavy quark are numerically significant. This will clarify 
the impact of heavy flavors on predictions relevant for future experiments such as FASER at the 
LHC, complementing the study in Ref.~\cite{Risse:2025smp}.

\acknowledgments
We are indebted to Stefano Forte and Felix Hekhorn for their interest and substantial support
throughout this work. We are grateful to Roman N.~Lee for assistance with his package
\texttt{Libra}~\cite{Lee:2020zfb}. We thank Marco Bonetti and Hai Tao Li for a careful reading of the manuscript and 
various conversations about the project. We also thank the following for helpful discussions: 
Konstantin Asteriadis, Arnd Behring, Daniel Baranowski, Long Chen, Maximilian Delto, Zhe 
Li and Jian Wang. This work is supported by the National Science Foundation of China under grants No.~12275156 and No.~12321005. 
Additionally, this work was started at the physics department of Milan University and partly 
supported by the European Research Council under the European Union’s Horizon 2020 research and 
innovation Programme (grant agreement n.740006).

\appendix
\addcontentsline{toc}{section}{Appendices} 
\section{Auxiliaries}\label{app:A}
\subsection{Parton kinematics}\label{app:kinematics}
Following Ref.~\cite{Aivazis:1993kh}, the standard DIS invariant variables are
\begin{equation}\label{eq:standardDISVars}
	P^2=M^2,\quad Q^2=-q^2,\quad x_B=\frac{Q^2}{2P\cdot q},
\end{equation}
where $P$ is the  momentum of the nucleon target, $q$ is the momentum transfer from the lepton 
vertex and $x_B$ is the Bjorken variable.

Using light-cone coordinates (cf. Ref.~\cite{Grozin:2005yg}), we parameterize the momenta of 
the off-shell boson and incident heavy quark in the DIS collinear frame as
\begin{equation}\label{eq:collinearFrameMomenta}
	\begin{aligned}
		q^{\mu} &\equiv (q_{1}^{+},q_{1}^{-};\vec{q}_{1}^{\perp}) = 
		\left(-\eta P^{+},\frac{Q^2}{2\eta P^+};\vec{0}\right), \\
		p^{\mu} &= \left(\xi P^{+}, 
		\frac{m^2}{2\xi P^+};\vec{0}\right),
	\end{aligned}
\end{equation}
respectively. Within this parameterization, the Bjorken variable generalizes to%
\begin{equation}\label{eq:genBjorkenVar}
	\frac{1}{\eta} = \frac{1}{2x_B} + \sqrt{\frac{1}{4x_B^2} + \frac{M^2}{Q^2}},
\end{equation}
where $\eta$ includes finite target-mass corrections.
	
The partonic center-of-mass energy relates to the momentum fraction $\xi$ through
\begin{equation}\label{eq:cmeParam}
	s = (q + p)^2 = \left(Q^2 + \frac{\eta}{\xi}m^2\right) %
	\left(\frac{\xi}{\eta} - 1\right).
\end{equation}
Inverting this equation results in
\begin{equation}\label{eq:xiDefinition}
	\xi = \eta \frac{Q^2 - m^2 + s + \Delta[-Q^2,m^2,s]}{2Q^2},
\end{equation}
where the triangle function is defined as usual:
\begin{equation}\label{eq:triagnleFunction}
	\Delta[a,b,c] = \sqrt{a^2 + b^2 + c^2 - 2(ab + ac + bc)}.
\end{equation}

Using Eq.~\eqref{eq:xiDefinition}, we obtain the lower bound of the integration domain
\begin{equation}\label{eq:prodThreshold}
	\xi_{\mathrm{th}} =%
	\begin{cases}
		\eta,& (s = 0) \\
		\eta\tfrac{1+\sqrt{1+4m^2/Q^2}}{2},& (s = m^2)\\[6pt]
		\eta \frac{Q^2 + 3m^2 + \Delta[-Q^2,\, m^2,\,4m^2]}{2Q^2}, & (s=4m^2)
	\end{cases}
\end{equation}
corresponding to the virtual and heavy-quark pair production thresholds, 
respectively. The upper bound follows from momentum conservation, giving the integration domain 
$\xi_{\mathrm{th}} \leq \xi \leq 1$ in Eq.~\eqref{eq:ansatzFactorization}.

In Ref.~\cite{Kretzer:1998ju}, a different scaling variable is introduced. Its definition reads
\begin{equation}\label{eq:varsKretzer}
	\xi' = \frac{\eta}{\xi}.
\end{equation}
This one is related to our choice of variables in Eq.~\eqref{eq:variableChange} via
\begin{equation}
	y_1 \to 1 - \xi',\; y_2 = m^2 \frac{\xi'}{Q^2 + m^2 \xi'}.
\end{equation}

\subsection[Projectors, normalization and $\gamma_5$ schemes]{Projectors, normalization and 
\texorpdfstring{$\bm{\gamma_5}$}{} schemes}\label{app:ProjANDNorm}
We work in dimensional regularization with $d = 4 - 2\epsilon$, using standard 
Clifford algebra relations. Special care is required 
when evaluating traces involving $\gamma_{5}$, which appears in the 
charged weak current defined in Eq.~\eqref{eq:chargedCurrentVector}.

The complication arises because $\gamma_{5}$ is intrinsically 
four-dimensional and cannot be consistently continued to $d$ dimensions: 
one cannot simultaneously preserve its anticommutation relation 
$\{\gamma_{\mu}, \gamma_5\} = 0$ and the cyclicity of traces involving 
$\gamma_5$ in $d$ dimensions~\cite{tHooft:1972tcz,Kreimer1990Clifford}.

Several schemes address this by compromising different algebraic properties. The 
Breitenlohner–Maison–'t Hooft–Veltman (HVBM) 
scheme~\cite{Breitenlohner:1975hg,Breitenlohner:1976te,Breitenlohner:1977hr} provides a 
consistent framework by abandoning the anticommutation relation $\{\gamma_\mu, \gamma_5\} = 
0$ for $\mu > 4$. In contrast, Kreimer's approach~\cite{Kreimer1990Clifford,Korner:1991sx} 
aims to preserve the cyclicity of traces but is less widely adopted in the community.

We employ two schemes to cross-check our results: Kreimer's and Larin's (a simplified HVBM 
variant)~\cite{Larin:1991tj,Larin1993Gamma5,Larin:1993tq}. We refer to the original literature 
for technical details and summarize only the essential projectors, normalization factors, and 
finite renormalizations relevant to our calculation.

Traces with $\gamma_5$ introduce Levi-Civita tensors. Their contraction 
gives
\begin{equation}\label{eq:LCSymbol}
	\varepsilon^{\mu_1\mu_2\mu_3\mu_4}\varepsilon_{\nu_1\nu_2\nu_3\nu_4} = 
	\det\begin{pmatrix}
		\delta^{\mu_1}_{\nu_1} & \cdots & \delta^{\mu_1}_{\nu_4} \\
		\vdots & \ddots & \vdots \\
		\delta^{\mu_4}_{\nu_1} & \cdots & \delta^{\mu_4}_{\nu_4}
	\end{pmatrix},
\end{equation}
where $\delta^{\mu_i}_{\nu_j}$ are Kronecker deltas, with reference 
dimension $d=4$ (Kreimer) or $d=4-2\eps$ (Larin).

The projectors derived from Eqs.~(\ref{eq:amplTensorStr},\;\ref{eq:extractFFs}) 
are
\begin{equation}\label{eq:projectors}
	\begin{aligned}
		\mathcal{P}_1^{\mu\nu} 
		&= \frac{1}{(2-d)}\Biggl(\eta^{\mu\nu}+
		\frac{1}{\Delta'}\Bigl\{-Q^2
		p^{\mu}_1p^{\nu}_1+m^2p^{\mu}_2p^{\nu}_2+
		(Q^2-m^2+s)(p^{\mu}_1p^{\nu}_2
		+p^{\mu}_2p^{\nu}_1)\Bigr\}\Biggr),\\[6pt]
		\mathcal{P}_2^{\mu\nu} 
		&= \frac{4}{(d-2)\Delta'}\Biggl\{-Q^2\eta^{\mu\nu}  + 
		\frac{2}{\Delta'}\Bigl[ 2 (d-1) 
		(Q^2)^2p_2^{\mu}p_2^{\nu} \\ &\qquad 
		+ \Bigl(\frac{1}{2}(d-2)(Q^2-m^2+s)^2-2m^2 Q^2\Bigr)p_1^{\mu}p_1^{\nu} \\
		&\qquad + (d-1) Q^2(Q^2-m^2+s) \bigl(p_1^{\nu}p_2^{\mu} + 
		p_1^{\mu}p_2^{\nu}\bigr)\Bigr]\Biggr\},\\[6pt]
		\mathcal{P}_3^{\mu\nu} 
		&= -\frac{4\mathrm{i}}{\Delta'}\,\varepsilon^{\mu\nu\alpha\beta}
		p_{1,\beta}\,p_{2,\alpha},
	\end{aligned}
\end{equation}
where the Levi–Civita tensor is taken to be strictly four-dimensional to derive the 
corresponding projectors and $\Delta'=\Delta[-Q^2,\,m^2,\,s]$. These projectors are directly 
applicable in Kreimer's scheme but require modifications for Larin's scheme.

Kreimer's scheme is implemented straightforwardly following 
Ref.~\cite{Korner:1991sx}. The main technical complication arises from scalar products of the 
form $\bar{\eta}_{\mu\nu}l_1^{\mu}l_2^{\nu}$, where the four-dimensional Minkowski 
tensor $\bar{\eta}_{\mu\nu}$ originates from contractions of pairs of LC tensors.

These can be evaluated either:
\begin{itemize}
	\item by Passarino-Veltman reduction \cite{Passarino:1978jh}, which expresses 
	$l_1^{\mu}l_2^{\nu}$ in terms of scalars and external tensors 
	before contracting with $\bar{\eta}_{\mu\nu}$; or
	\item via generating functions from Symanzik polynomials as detailed in 
	Refs~\cite{Bern:2002tk,Heller:2020owb}, where $-2\eps$-dimensional scalar products are 
	replaced with 
	combinations of inverse propagators.
\end{itemize}
We used the first method for all such scalar products. Where applicable, the second method 
provided a valuable cross‑check of the results obtained with the first.

We implement Larin's scheme following Ref.~\cite{Moch2015Gamma5}. Spurious 
coefficients from products of Levi–Civita tensors in $d$ dimensions are corrected via
\begin{equation}
	\begin{aligned}
		F_1 &: A^2 \to \frac{6}{(d-2)(d-3)(d-7)}A^2,\\[4pt]
		F_2 &: A^2 \to -\frac{2}{(d-2)(d-3)}A^2,\\[4pt]
		F_3 &: VA \to \frac{1}{(d-2)(d-3)}VA,
	\end{aligned}
\end{equation}
where $V$ and $A$ denote vector and axial parts of the vertex (cf. 
Eq.~\eqref{eq:chargedCurrentVector}).

To restore the axial Ward identity, we apply finite renormalization to the 
axial current $A\to Z_A Z_5A$ \cite{Larin:1991tj}
\begin{equation}\label{eq:finiteRenormLarin}
	\begin{aligned}
		Z_A &= 1 + \frac{C_F}{2\eps}\left(\frac{\alpha_s}{2\pi}\right)^2
		\left(\frac{11C_A - 6n_f}{3}\right) + \mathcal{O}(\alpha_s^3),\\
		Z_5 &= 1 - 2C_F\left(\frac{\alpha_s}{2\pi}\right) \\ &\quad + 
		\left(\frac{\alpha_s}{2\pi}\right)^2\left(\frac{11}{2}C_F^2 - \frac{107}{36}C_F C_A + 
		\frac{1}{18}C_F n_f\right) + \mathcal{O}(\alpha_s^3).
	\end{aligned}
\end{equation}
\section{Master integrals material}\label{app:B}
\paragraph{Alphabets.}
The relevant alphabets appearing in Eq.~\eqref{eq:fuchsForm} are listed below. The remaining 
alphabets from that equation are not used and are therefore omitted.
\begin{align}
	\nonumber \{a^{\mathrm{I}}_{2,i}\} =& \left\{0,\,-1,\,-\frac{1}{2},\, 1-y_1,\, \frac{1}{2}
	\left(1-y_1\right),\, -\frac{y_1}{4},\, -\frac{1}{y_1},\, \frac{1}{y_1},\right.\\ 
	&\left.\frac{1}{y_1 - 4},\, \frac{1}{y_1 - 3},\, \frac{1-y_1}{y_1},\, -\frac{y_1}{y_1+1},\, 
	-\frac{y_1}{y_1+2}\right\}, \label{eq:ab_1} \\
	\{a^{Q}_{2,i}\} =& \left\{0,\,\frac{1-y_1}{y_1}\right\}, \label{eq:ab_2} \\%
	\{a^{\mathrm{II}}_{2,i}\}=&\left\{0,1,2-y_1,\frac{1}{y_1},y_1,\frac{y_1}{2 y_1-1}\right\}, 
	\label{eq:ab_3}\\%
	\{a^{\mathrm{III}}_{1,i}\} =& \left\{0,\,-1,\,-\frac{1}{y_2},\,\frac{1}{y_2},\,\frac{1 - 
		y_2}{y_2},\,\frac{1}{y_2 - 1},\,\frac{2}{y_2 - 1},\,\frac{1}{2y_2 - 1}\right\}, 
	\label{eq:ab_4}.
\end{align}
The alphabets listed below appear in the systems for boundary conditions as in 
Eq.\eqref{eq:boundaryEqs}.
\begin{align}
	\{b^{\mathrm{I}}_i\}=&\{0,\,1,\,2\}\label{eq:ab_5},\\
	\{b^{\mathrm{II}}_i\}=&\{0,\,1,\,2\}\label{eq:ab_6},\\
	\{b^{\mathrm{III}}_i\}=&\{0,\,1\}\label{eq:ab_7},\\
	\{b^{\mathrm{IV}}_{i}\} =& \{0, 1, -1, 1/2\}\label{eq:ab_V},\\
	\nonumber \{a^{u}\} =&%
	\nonumber \left\{0,\,-1,\,-2,\,-\frac{2}{2-y_1},\,-\frac{2\left(1-y_1\right)}{2-y_1},\,%
	\frac{y_1}{r_1}-1,\,-\frac{y_1}{r_1}-1,\,\frac{y_1}{r_2}-1,%
	\right.\\ &\left.%
	\nonumber -\frac{y_1}{r_2}-1,\, \frac{y_1}{r_3}-1,\, -\frac{y_1}{r_3}-1,\, 
	\frac{r_3+y_1-2}{2-y_1},%
	\frac{r_3-y_1+2}{y_1-2},\, \frac{y_1}{r_4}-1,%
	\right.\\&\left.%
	\nonumber-\frac{y_1}{r_4}-1,\, r_5-1,\, -r_5-1,\, r_6-1,\, -r_6-1,\, \frac{y_1}{r_7}-1,\, 
	-\frac{y_1}{r_7}-1,\,
	\right. \\ &\left.%
	\frac{y_1}{r_8}-1,\, -\frac{y_1}{r_8}-1,\, r_9-1,\, -r_9-1\right\} \label{eq:ab_8}.
\end{align}
The simplifications in the integrand (Eq.~\eqref{eq:goncharovIterIntDef}) produce multiple 
square roots due to the transformation defined in Eq.~\eqref{eq:squareRoot} as can be seen in 
the sets above. We collect these below
\begin{equation}
	\begin{aligned}
		r_1 =& \sqrt{\left(2-y_1\right)y_1},\,%
		r_2 = \sqrt{\left(4-3 y_1\right)y_1},\,%
		r_3 = \sqrt{\left(y_1-4\right)y_1},\\%
		r_4 =& \sqrt{\left(y_1-2\right)y_1},\,%
		r_5 = \sqrt{\frac{y_1+1}{y_1-3}},\,%
		r_6 = \sqrt{\frac{y_1+2}{y_1-2}},\\%
		r_7 =& \sqrt{y_1-4},\,%
		r_8 = \sqrt{y_1+4},\,%
		r_9 = \sqrt{\frac{\left(y_1-3\right) y_1}{y_1^2-3 y_1+4}}.
	\end{aligned}
\end{equation}
Here, when working with the square roots, it is important to keep in mind the Feynman 
prescription which can be easily inferred via $s\to s + \mathrm{i}0$.

\paragraph{Cut mappings.} To obtain the cut integral families, we use the rules provided in 
Tab.~\ref{tab:cutRules}. The rules are specific to our setup; for instance, families `fam2' and 
`fam3' contain multiple cuts of the same type (e.g., $\mathrm{II}'$). The cut families listed 
above are not all independent, i.e. some can be mapped onto others, resulting in a smaller set 
of unique families. In our practical computation, only the cut rules for `fam1' are used 
directly.
\begin{table}[h]
	\centering
	\begin{tabular}{c|c|c|c|c|c|c}
		\toprule
		Cut & fam1 & fam2 & fam3 & fam4 & fam5 & fam6 \\
		\hline
		$\mathrm{I}'$ & $\{1, 3, 7\}$ & -- & -- & $\{2, 3, 5\}$ & $\{2, 3, 5\}$  & -- \\
		$\mathrm{II}'$ & -- & -- & -- & --  & -- & $\{1, 3, 5\}$ \\
		$\mathrm{III}'$ & $\{2, 4, 6\}$ & $\{1, 4, 6\}$ & $\{1, 4, 6\}$ & -- & --   & --\\
		$\mathrm{IV}'$ & $\{6, 7\}$ & $\{3, 4\}$ & $\{3, 4\}$ & -- & -- & -- \\
		\bottomrule
	\end{tabular}
	\caption{On-shell propagator assignments for each cut and generic integral family. The 
	entries list the indices $i$ of the inverse propagators $D_i$ (defined in 
	Tab.~\ref{tab:integralFamilies}) to be put on-shell according to 
	Eq.~\eqref{eq:onShellProps}.}
	\label{tab:cutRules}
\end{table} 
\paragraph{Normalization.}
The factors $f_i$ are
\begin{equation}\label{eq:normalizationFactor}
	\begin{aligned}
		f_{\mathrm{I}} &= \tfrac{(16\pi^2)^2}{2\pi} \cdot 
		(2 - 13\epsilon + 27\epsilon^2 - 18\epsilon^3),\\
		f_{\mathrm{II}} &= \tfrac{(16\pi^2)^2}{\pi} \cdot 
		\epsilon\frac{-2 + 13\epsilon -27\epsilon^2 + 18 \epsilon^3}{2(\epsilon - 1)},\\
		f_{\mathrm{III}} &= 2\tfrac{(16\pi^2)^2}{\pi} \cdot 
		\epsilon(1 - 3\epsilon + 2\epsilon^2),\\
		f_{\mathrm{IV}} &= 2\tfrac{(16\pi^2)^2}{2} \cdot 
		\epsilon^2(2 - 7\epsilon + 6\epsilon^2).
	\end{aligned}
\end{equation}
The overall factor of 2 in $f_{\mathrm{III}}$ and $f_{\mathrm{IV}}$ originates from the optical 
theorem (Eq.~\eqref{eq:opticalTheorem2}); remaining prefactors are chosen  to obtain the UT 
form of master integral and for general convenience.

\bibliographystyle{JHEP}
\bibliography{biblio.bib}

\end{document}